\documentstyle[12pt,a4]{article}
\begin{document}
\thispagestyle{empty}
\vspace{1.5cm}
DESY 92-071
\begin{center}
{\large \bf BILOCAL FIELD APPROACH AND SEMILEPTONIC HEAVY MESON DECAYS}
\vspace{1.5cm}

\renewcommand{\thefootnote}{\fnsymbol{footnote}}

{\large Yu. L. Kalinovsky} \footnote[1]{e-mail address:
KALINOVS@THEOR.JINRC.DUBNA.SU}
\vspace{0.5cm}

{\em Joint Institute for
Nuclear Research, Dubna \\
Head Post Office, P.O. Box 79, 101000 Moscow, Russia}
\vspace{0.5cm}

{\large L. Kaschluhn} \footnote[7]{e-mail address: IFHKAS@DHHDESY3.BITNET}
\vspace{0.5cm}

{\em DESY -- Institut f\"ur Hochenergiephysik \\Platanenallee 6,
O--1615 Zeuthen, Germany}
\end{center}

\vspace{4.5cm}
\begin{abstract}
{\normalsize
In this paper we consider the bilocal field approach for $QCD$.
We obtain a bilocal effective meson action with a potential kernel given in
relativistic covariant form. The corresponding Schwinger--Dyson and
Bethe--Salpeter equations are investigated in detail. By introducing
weak interactions into the theory we study heavy meson properties as
decay constants and semileptonic decay amplitudes.
Thereby, the transition from the bilocal
field description to the heavy quark effective theory is discussed.
Considering as example the semileptonic decay of a pseudoscalar $B$--meson
into a pseudoscalar $D$--meson we obtain an integral expression for the
corresponding Isgur--Wise function in
terms of meson wave functions.}
\end{abstract}

\renewcommand{\thefootnote}{\arabic{footnote}}

\newpage

\vspace{3.0cm}

{\large \bf 1. Introduction }

\vspace{0.5cm}

The investigation of heavy meson decays has become one of the important
problems in heavy flavour physics. Especially, $ B $--decays $[1]$ play an
important role in determining the Kobayashi--Maskawa matrix elements,
including the $ CP $ violating phase. Furthermore, rare decays of heavy
mesons may indicate deviations from the standard model.
For the description of physical (hadronic) decay processes one needs to know
the wave functions and form factors. There exist several approaches to
attack these problems. Examples are phenomenological quark models, $ QCD $
sum rules, heavy quark effective theory, potential models.

Here we shall consider one more possibility -- potential models in the bilocal
field approach for $ QCD $ $[2,3]$.
Thereby, we start from an approximate $ QCD $ action with massive quark fields
and hadronize it.
Substitution of the interaction kernel containing the free gluon propagator
by a instantaneous Lorentz--vector potential kernel yields a class of bilocal
potential models for meson fields $[4,5]$. Potential models do not possess
full relativistic invariance because the Fock space is restricted to
$(q \bar{q})$ pairs and an instantaneous interaction is assumed. Nevertheless,
in our model the kernel is written
in relativistic covariant form by introducing a special vector being
proportional to the bound--state total momentum. This allows us to investigate
equations for a bound state moving together with the potential kernel.
This fact is important especially for the calculation of formfactors where one
needs to know meson wave functions for moving particles. One further
advantage of our relativistic model consists in the possibility of
describing in a uniform manner both light and heavy mesons in dependence on
the choice of the potential. So, in the case of a short--range potential one
can use the separable approximation (for small orbital momenta) to obtain a
new regularized version of the Nambu--Jona--Lasinio model $[6]$. We should
also mention that we are able to extend some of the results obtained by
A. Le Yaouanc et al. $[7]$, who investigated a local quark model for massless
quark fields within the Hamilton formalism with an instantaneous
fourth--component Lorentz--vector colour confining potential (e.g. the
harmonic oscillator potential).

In the first part of this paper the Schwinger--Dyson and Bethe--Salpeter
equations for our model will be derived,
and the meson wave functions fulfilling Schr\"odinger--like equations will be
introduced. Thereby, the equations for the quark mass spectrum will be
considered in application to oscillator, Coulomb and linear potentials.
Concerning the equations on bound state masses special attention will be
paid to the derivation of equations for wave functions of scalar, pseudoscalar
and vector particles.

Then, in the second part of the paper we will apply the bilocal field method
to heavy meson physics. Therefore we will introduce semileptonic weak
interactions into the underlying potential theory by
shifting the bilocal field correspondingly. Furthermore, now we will have to
consider in general moving bound states. We will derive formulas for
pseudoscalar meson decay constants as well as for semileptonic decays of heavy
quarkonia. Moreover, we will establish a relation between bilocal field
approach and heavy quark effective theory. This will allow us to consider the
semileptonic decays in the limit to the heavy quark effective theory
$[8-10]$. Thereby, as example the Isgur--Wise function $[8]$ corresponding to
decays $B \rightarrow D(l \nu_{l})$ will be defined within the bilocal field
method.

In this paper general formulas containing the so far undetermined meson
wave functions will be obtained. Nevertheless, as will be shown, these wave
functions fulfil systems of integral equations which shall be solved in the
near future for concrete potentials by different methods.

The paper is organized as follows. In sect. 2 the formulation of the model is
given. The corresponding equations for the quark spectrum within a meson are
derived in sect. 3. Sect. 4 contains the investigation of equations for
bound state vertex and wave functions. In sect. 5 formulas for the
pseudoscalar meson decay constants are derived. In sect. 6 we discuss the
relation between bilocal field approach and heavy quark effective theory.
Semileptonic decays of heavy
quarkonia in the limit of heavy quark effective theory are investigated in
sect. 7. The summary and conclusions are given in sect. 8.

\vspace{0.5cm}
{\large \bf 2. Formulation of the model }
\vspace{0.5cm}

{\bf 2.1. Hadronization of QCD}
\vspace{0.5cm}

Let us start with the approximate QCD action
for quarks $q$ in the form
\begin{equation}
W = \int  d^{4}x \bar{q}(x) [ G_0^{-1} (x) ] q(x) -
\frac{g^{2}}{2} \int \int d^{4}x d^{4}y j^{a}_{\mu}(x) D^{ab}_{\mu\nu}(x-y)
j^{b}_{\nu}(y) \,\,\, .
\end{equation}
Here $G_0$ is the Green's function for free quarks,
\begin{eqnarray*}
G_0^{-1} = i \rlap/\partial - \hat{m}_0 \,\, ,
\end{eqnarray*}
where $\hat{m}_0 $ is the bare quark mass matrix,
$ \hat{m}_0 = \mbox{diag}(m^0_1, m^0_2, \ldots , m^0_{N_f}) $,
$N_f$ being the flavour number. The quark current
$j^{a}_{\mu}(x)$ is defined by the relation
\begin{eqnarray*}
j^a_{\mu} (x) = \bar{q} (x)
\biggl( \frac{\lambda^a}{2} \biggr) \gamma_{\mu} q (x) \,\,,
\end{eqnarray*}
where $\lambda^a$
are the Gell--Mann matrices in colour space $\mbox{SU(3)}_c $,
$\mu$ is the Lorentz index and $\gamma_{\mu}$ denotes the
Dirac matrix. The quark--gluon interaction with coupling constant $ g $ is
mediated via the free gluon propagator
$ D^{ab}_{\mu\nu}(x-y) \equiv \delta^{ab} g_{\mu\nu} D(x-y) $.

For hadronization of action (1) let us first rewrite the bilocal
interaction term
\begin{eqnarray*}
- \int \int d^{4}x d^{4}y \frac{1}{2}
\biggl( \bar{q}(x) \frac {\lambda^{a}} {2} \gamma_\mu q(x) \biggr)
\frac{g^{2}}{2} D(x-y)
\biggl( \bar{q}(y) \frac {\lambda^{a}} {2} \gamma^\mu q(y) \biggr)
\end{eqnarray*}
in the form
\begin {equation}
\int \int d^{4}x d^{4}y  q_{B}(y) \bar{q}_{A}(x) K_{AB,EF}(x-y)
q_{F}(x) \bar{q}_{E}(y)
\end {equation}
with the kernel
\begin{eqnarray*}
K_{AB,EF}(x-y) = {\gamma}_{ru}^{\mu}(\gamma_{\mu})_{ts}
\sum_{a=1}^{8} {\frac {\lambda_{\alpha \delta}^{a}} {2}}
{\frac {\lambda_{\gamma \beta} ^{a}} {2}}
\delta_{il} \delta_{kj}
\frac{g^{2}}{2}  D(x-y) \,\,\, .
\end {eqnarray*}
Here $ A, B, E, F $ are short--hand notations for the indices
$ A = \{r,\alpha,i\}, \ B = \{s,\beta,j\}, \\
E = \{t,\gamma,k\} $ and
$ F = \{u,\delta,l\} $ \,\,. The first index in the bracket refers to the
Lorentz group , the second one to the colour group $ SU(3)_{c} $ and the
third one to the flavour group $ SU(N_{f})_{f} $.

Now, we make the colour Fierz rearrangement $[11]$
\begin{eqnarray}
\sum_{a=1}^{8}
\lambda^{a}_{\alpha \delta}
\lambda^{a}_{\gamma \beta} = \frac{4}{3}
\delta_{\alpha \beta} \delta_{\gamma \delta}
+ \frac{2}{3} \sum_{\rho=1}^{3}
\epsilon_{\rho \alpha \gamma}  \epsilon^{\rho \beta \delta }
\end{eqnarray}
with $ \epsilon_{ \alpha \beta \gamma} $ being the
antisymmetric Levi- Civita tensor. This identity allows to decompose
$ K_{AB,EF}(x-y) $ completely into "attractive" colour singlet
($ q \bar{q} $) and colour antitriplet ($q q$) channels.
In this treatment "repulsive" colour octet  ($ q \bar{q} $)  and
sextet ($q q $) channels are absent in a natural way.
In this paper we want to discuss $ (q \bar{q}) $ meson bound states.
For this reason we will consider only the colour singlet part of (3) and
therefore the kernel
\begin {equation}
K_{AB,EF}^{singlet}(x-y) = {\gamma}_{ru}^{\mu}(\gamma_{\mu})_{ts}
\frac{1}{N_{c}} \delta_{\alpha\beta} \delta_{\gamma\delta}
\delta_{il} \delta_{kj}
\frac{g^{2}}{2}  D(x-y) \,\,,
\end {equation}
where $ N_{c} = 3 $ denotes the colour number.

Inserting the kernel (4) into the bilocal interaction term (2)
one obtains (omitting the group indices)
\begin{eqnarray*}
\frac{1}{N_{c}} \int \int d^{4}x d^{4}y  q(y) \bar{q}(x) \gamma^{\mu}
\frac{g^{2}}{2} D(x-y) q(x) \bar{q}(y) \gamma_{\mu} \,\,.
\end {eqnarray*}
Then, action (1) can be represented in the form
\begin{eqnarray*}
W_{\em M} &=& \int \int d^{4}x d^{4}y
\biggl\{
\bigl( q(y) \bar{q}(x) \bigr)
\bigl( - G_0^{-1} \bigr) \delta (x-y) \\
&& \,\,\,\,\,\,\,\,\,\, + \frac{1}{2N_{c}}
\Bigl[
\bigl( q(y) \bar{q}(x) \bigr) {\cal{K}} (x-y) \bigl( q(x) \bar{q}(y) \bigr)
\Bigr]
\biggr\} \,\,
\end{eqnarray*}
with the Lorentz--vector kernel for $QCD$
\begin{eqnarray}
{\cal{K}}(x-y) = g^{2} \gamma^{\mu} D(x-y) \otimes \gamma_{\mu} \,\,.
\end{eqnarray}
In symbolic notation it reads
\begin{eqnarray*}
W_{\em M} =
\bigl( q \bar{q}, - G_0^{-1}  \bigr) + \frac{1}{2N_{c}}
\bigl(q \bar{q}, {\cal{K}} \, q \bar{q} \bigr) \,\, .
\end{eqnarray*}

\vspace{0.5cm}

{\bf 2.2. Schwinger--Dyson and Bethe--Salpeter equations}
\vspace{0.5cm}

Let us now consider the functional integral
\begin{eqnarray*}
Z = \int {\cal D}q {\cal D}\bar{q}
\mbox{exp} \bigl\{ i W_{\em M} [q, \bar{q} ]    \bigr\} \,\, .
\end{eqnarray*}

After integrating over the quark fields with the help of the
Legendre transform one gets
\begin{eqnarray*}
Z = \int {\cal D} {\cal M}
\mbox{exp} \biggl\{ i W_{eff} [ {\cal M} ]    \biggr\} \,\,\, ,
\end{eqnarray*}
with the effective action
\begin{eqnarray}
W_{eff} [ {\cal M} ] =
N_c
\biggl\{
- \frac{1}{2} ( {\cal M}, {\cal{K}}^{-1} {\cal M} )
-i \mbox{Tr} \mbox{ln} ( -G_0^{-1} + {\cal M} )
\biggr\}
\end{eqnarray}
for meson fields $ {\cal{M}} $.
Here $\mbox{Tr}$ means integration over the continuous variables
and taking the traces over the discret ones (spinor and flavour indices).
The condition of minimum for this effective action reads
\begin{eqnarray}
{\cal{K}}^{-1}{\cal M} +i \frac{1}{-G_0^{-1}+{\cal M} } = 0 \,\, .
\end{eqnarray}
Let us denote the solution of this equation by
($\Sigma - \hat{m}^0 $). Then one obtains from (7) the
Schwinger--Dyson equation
\begin{eqnarray}
\Sigma = \hat{m}^0 + i {\cal{K}} G_{\Sigma} \,\, ,
\end{eqnarray}
where
\begin{eqnarray*}
G^{-1}_{\Sigma} = i \rlap/\partial - \Sigma \,\, .
\end{eqnarray*}

Expanding action (6) around the minimum with
$ {\cal M} = (\Sigma - \hat{m}^0 ) +  {\cal M}^{\prime}  $ one
gets
\begin{eqnarray}
W_{eff} [{\cal M}] &=& W_{eff} [ \Sigma ] \nonumber \\
&+& N_{c} \biggl\{
- \frac{1}{2} ( {\cal M}^{\prime}, {\cal{K}}^{-1}
{\cal M}^{\prime} ) - \frac{i}{2}
{\rm{Tr}} ( G_{\Sigma} {\cal M}^{\prime} )^2
- i \sum^{\infty}_{n=3} \frac{1}{n} {\rm{Tr}}
(- G_{\Sigma} {\cal M}^{\prime} )^n
\biggr\} \,\, .
\end{eqnarray}
The vanishing of the second variation of this effective action
with respect to $ {\cal M}^{\prime}  $ ,
\begin{eqnarray*}
\frac{\delta^2 W_{eff}}{ \delta {\cal M}^{\prime}  \delta {\cal M}^{\prime} }
{\Bigg \vert}_{ {\cal M}^{\prime} =0 }
\cdot \Gamma = 0 \,\, ,
\end{eqnarray*}
leads to the Bethe--Salpeter equation
\begin{eqnarray}
\Gamma
= -i {\cal{K}} (G_{\Sigma}  \Gamma  G_{\Sigma})
\end{eqnarray}
for the vertex function $\Gamma (x,y)$ in the ladder approximation.

The Schwinger--Dyson equation (8) describes the quark spectrum
in the meson, whereas the Bethe--Salpeter equation (10)
yields the bound state spectrum. Solving together both
equations one may obtain the wave functions of the bound
states and calculate with their help not only static properties
of mesons as the mass spectrum and the decay constants but also
decay probabilities.

\newpage

{\bf 2.3. Relativistic covariant description of potential models }
\vspace{0.5cm}

{}From the translation invariance of the two--particle bound
state it follows that one can separate the c.m. motion of the
($\bar{q} q$) system from the relative motion. But
in a relativistic theory it is impossible to separate the c.m.
coordinates. Nevertheless, one may separate the total momentum
${\cal P}_{\mu} $ being the momentum of the c.m. motion.
After this one can more or less arbitrarily define some coordinate
$X_{\mu}$ representing the absolute position in space--time and use the
relative coordinate of the bound state only. For example, let
$X_{\mu}$ be the position $x$ or $y$ of one of the particles
(quarks) or a linear combination
$X =  \alpha x + (1- \alpha )y  $ of them. For $\alpha = 1$ the relative
coordinate is $z= x-y$. In this case one can look for the solution
of the Bethe--Salpeter equation for the bound state wave function
$\psi$ in the form
\begin{eqnarray*}
\psi (x,y) =  e^{ i{\cal P}X } \psi (z) \,\,\, .
\end{eqnarray*}

Let us now substitute the $ QCD $ kernel
$ {\cal{K}} $ of the integral equations (8) and (10) by a
potential one. This can be done in a relativistic covariant way by
choosing instead of (5)
\begin{eqnarray}
{\cal{K}}^{\eta}(x,y) = {\cal{K}}^{\eta}(x-y \vert \frac{x+y}{2}) =
\rlap/\eta
V( z^{\perp} ) \delta ( z \cdot \eta ) \otimes \rlap/\eta
\end{eqnarray}
with
\begin{eqnarray*}
z_{\mu}^{\perp} = z_{\mu} - z_{\mu}^{\vert \vert }
\,\,\, , \,\,\,  z_{\mu}^{\vert \vert } = \eta_{\mu} ( z \cdot \eta )
\,\,\, .
\end{eqnarray*}
Here $\eta_{\mu}$ is a vector
($ \rlap/\eta = \eta_{\mu} \gamma_{\mu}; \eta^2 = 1 $), being proportional
to the momentum eigenvector ${\cal P}_{\mu}$ and  describing the motion of
the bound state as a whole
\begin{eqnarray*}
\eta_{\mu} = \frac{{\cal P}_{\mu} }{\sqrt{{\cal P}^2}} \,\,\, ,
\,\,\,
{\cal P}_{\mu} \psi = -i
\frac{\partial }{\partial X_\mu } \psi \,\,\, .
\end{eqnarray*}
In (11) the transversality of the exchange interaction in the ($ q\bar{q} $)
system is ensured by $ V( z^{\perp} ) $ -- some phenomenological potential
for the description of quarkonia. Furthermore,
the $ \delta-$function $ \delta ( z \cdot \eta ) $ guarantees the
instantaneousness of the exchange interaction.

We should add that one can arrive at equation (11) by discussing moving
bound states rigorously within the quantization theory for gauge theories
$ [12] $.
For the bound state
at rest one has  $\vec{\cal P}= 0  $, so that
$ \eta_{\mu} = (1,0,0,0)  $ and the kernel takes the form
\begin{eqnarray}
{\cal{K}}^{\eta} =  \gamma_0 V(z^{\perp}) \delta (z_0) \otimes \gamma_0 \,\,.
\end{eqnarray}
Kernel (12) is known from the calculation of the ($e^+ e^-$)
positronium spectrum in electrodynamics with
$V$ as Coulomb potential, where an exchange of transversal photons
takes place.

Phenomenologically we may choose the interaction potential in
the form
\begin{eqnarray}
V(r) = - \frac{4}{3} \frac{\alpha_s }{r} +
V_0 \delta (r) + a r + b r^2  \,\, .
\end{eqnarray}
Here the first term is a Coulomb--type potential for one--gluon
exchange, whereas the last two terms in (13) guarantee quark confinement.
The second term leads to the Nambu--Jona--Lasinio model which
has been considered within this approach in $[6]$.

The potential (13) may be attributed to the sum of the Coulomb
and oscillator potentials. In the applications we will consider
only two potentials $V(r)$, the first one being the sum
of Coulomb, constant and linear potentials and the second one --
the sum of Coulomb, constant and oscillator potentials.
In Table 1. all potentials under consideration are displayed in
$x$ space as well as in momentum space. (There the definition
$ \Delta_{ {\bf p} } = \partial^2 / \partial {\bf p}^2 $ has
been used.)

\vspace{0.5cm}

{\large \bf 3. Equations for the quark mass spectrum within a meson}

\vspace{0.5cm}

{\bf 3.1. The case of an unspecified potential }
\vspace{0.5cm}

The Schwinger--Dyson equation (8) takes for the potential kernel (11) in
momentum space the form
\begin{eqnarray}
\Sigma({\bf{p}}) = \hat{m}^0 + i \int \frac{d^{4}q}{(2 \pi)^4}
V({\bf p} - {\bf q})
\rlap/\eta G_{\Sigma} (q) \rlap/\eta \,\, ,
\end{eqnarray}
where
\begin{eqnarray*}
V({\bf p} - {\bf q}) &=& \int d^{4}x e^{-i (p-q)x} V(x^{\perp})
\delta (x \cdot \eta) \,\,\, , \\
G_{\Sigma}(q) &=& \int d^{4}x e^{-i qx} G_{\Sigma}(x) \,\,\, .
\end{eqnarray*}

Assuming flavour diagonality of $ \Sigma({\bf p}) $, i.e. considering
$ \Sigma = {\rm{diag}} (\Sigma_{1}, \Sigma_{2}, ..., \Sigma_{N_{f}}) \,,$
equation (14) falls into identical equations for all $ \Sigma_{n} $ with
bare mass term $ m^{0}_{n} $ , $ n=1,2, ...,N_{f} $. Omitting the flavour
index $ n $, these Schwinger--Dyson equations for the fields $ \Sigma $ of a
given flavour read in the rest frame, $ \eta_{\mu} = (1,0,0,0) $:
\begin{eqnarray}
\Sigma({\bf{p}}) = m^0 + i \int \frac{d^{4}q}{(2 \pi)^4}
V({\bf p} - {\bf q})
\gamma_{0} G_{\Sigma} (q) \gamma_{0} \,\, .
\end{eqnarray}
Now we make the ansatz
\begin{eqnarray}
\Sigma({\bf p}) = A({ \bf p}) \vert {\bf p} \vert + B({\bf p}) p_{i}
\gamma_{i} \,\,,\,\,\,\,\,\,\,\,\,\, i=1,2,3 \,.
\end{eqnarray}
Then, the propagator $ G_{\Sigma}(q) $ for the free quark
can be represented as follows
\begin{eqnarray}
G_{\Sigma}(q) =
\frac{1}{\rlap/{q} - \Sigma({\bf q}) + i\epsilon}  &=&
\biggl(
\frac{\Lambda_+({\bf q})}{ q_0 -E({\bf q}) +i \epsilon } +
\frac{\Lambda_-({\bf q})}{ q_0 +E({\bf q}) -i \epsilon }
\biggr) \gamma_0    \nonumber \\ \\
&=& \gamma_0 \biggl(
\frac{\bar{\Lambda}_+({\bf q})}{ q_0 -E({\bf q}) +i \epsilon } +
\frac{\bar{\Lambda}_-({\bf q})}{ q_0 +E({\bf q}) -i \epsilon }
\biggr) \,\,\, . \nonumber
\end{eqnarray}
Here the notations
\begin{eqnarray}
\Lambda_{\pm}({\bf q}) &=&
S^{-1}({\bf q}) \Lambda_{\pm}^0 S({\bf q}) =
\frac{1}{2} ( 1 \pm S^{-2}({\bf q}) \gamma_0 ) =
\frac{1}{2} ( 1 \pm \gamma_0 S^{2}({\bf q})  ) \,\,\, ,
\nonumber \\ \\
\bar{\Lambda}_{\pm}({\bf q}) &=&
S({\bf q}) \Lambda_{\pm}^0 S^{-1}({\bf q}) =
\frac{1}{2} ( 1 \pm S^{2}({\bf q}) \gamma_0 ) =
\frac{1}{2} ( 1 \pm \gamma_0 S^{-2}({\bf q})  ) \nonumber
\end{eqnarray}
and
\begin{eqnarray}
E({\bf q}) = \vert {\bf{q}} \vert \bigl[ A^{2}({\bf q})
+ (1+B({\bf q}))^{2} \bigr]^{1/2}
\end{eqnarray}
have been introduced, where
\begin{eqnarray}
\left.
\begin{array}{rl}
S^{\pm 2}({\bf q}) =& \mbox{sin} \phi ({\bf q})  \pm \hat{\bf q} \mbox{cos}
\phi ({\bf q}) =
\mbox{exp} \{ \pm 2 \hat{\bf q} \nu ({\bf q})  \} \,\, , \\ \\
\mbox{sin} \phi ({\bf q}) =& {\displaystyle \frac{A({\bf q}) \vert {\bf q}
\vert} {E({\bf q})}  }\,\, , \,\,\, \,\,\,\,\,\,\,\,\,
\mbox{cos} \phi({\bf q}) = {\displaystyle \frac{ (1+B({\bf q})) \vert {\bf q}
\vert }{E({\bf q})}  }\,\,, \\ \\
S^{\pm 1}({\bf q}) =& \mbox{cos} \nu({\bf q})  \pm \hat{\bf q} \mbox{sin}
\nu ({\bf q})\,\, , \,\,\,\,\,\,\,\,
\nu ({\bf q}) = \frac{1}{2} ( - \phi ({\bf q})+ \frac{\pi}{2} ) \,\, , \\ \\
\hat{\bf q} =& \hat{q}_i \gamma_i \,\, , \,\,\,\,\,\,
\hat{q}_i  = {\displaystyle \frac{q_i}{ \vert {\bf{q}} \vert}}\,\, , \,\,\,\,
\,\, \hat{\bf q}^2 = -1 \,\,\, ,
\end{array}
\right \}
\end{eqnarray}
and
\begin{eqnarray}
\Lambda^0_{\pm} = \frac{1}{2} ( 1 \pm \gamma_0 ) \,\, .
\end{eqnarray}

With these definitions one can write the solution $ \Sigma({\bf p}) $ of the
Schwinger--Dyson equation (15) in the form
\begin{eqnarray}
\Sigma ({\bf p}) =
E({\bf p}) \mbox{sin}\phi ({\bf p}) +
\hat{\bf p} ( E({\bf p}) \mbox{cos}\phi ({\bf p})- \vert {\bf p} \vert )
= ( {\bf p} {\bf\gamma}  ) + E({\bf p}) S^{-2}({\bf p})  \,\, .
\end{eqnarray}
By inserting (17) and (22) into (15) we can rewrite the Schwinger--Dyson
equation as
\begin{eqnarray*}
E({\bf p}) \mbox{sin} \phi ({\bf p}) &-&
\hat{\bf p} ( E({\bf p}) \mbox{cos}\phi ({\bf p}) - \vert {\bf p} \vert )  \\
&=&  m^0 - \frac{1}{2} \int \frac{d {\bf q}}{(2 \pi)^3} V({\bf p}-{\bf q})
\gamma_0 \bigl( 1 - \gamma_0 (
\mbox{sin}\phi({\bf q}) + \hat{\bf q}\mbox{cos}\phi({\bf q}) )
\bigr) \,\, .
\end{eqnarray*}
Taking the trace on both sides of this equation, one obtains a
system of two equations,
\begin{eqnarray}
\left \{ \begin {array} {ll}
E({\bf p}) \mbox{sin}\phi ({\bf p}) = m^0 + {\displaystyle \frac{1}{2}
\int \frac{d {\bf q}}{(2 \pi)^3}}
V ( {\bf p} - {\bf q}) \mbox{sin}\phi ({\bf q}) \\
E({\bf p}) \mbox{cos}\phi ({\bf p}) =
\vert {\bf p} \vert
 + {\displaystyle \frac{1}{2} \int \frac{d {\bf q}}{(2 \pi)^3} }
V ( {\bf p} - {\bf q}) \chi({\bf p},{\bf q}) \mbox{cos}
\phi ({\bf q}) \,\, ,
\end{array}
\right.
\end{eqnarray}
which defines the mass spectrum of two quarks forming a bound state.
Here the notation
\begin{eqnarray}
\chi({\bf p},{\bf q})= (\hat{\bf q} \cdot \hat{\bf p})
= \mbox{cos}\theta ({\bf p},{\bf q})
\end{eqnarray}
has been introduced.
For solving the system of equations (23) it is necessary to fix
the interaction potential.

\vspace{0.5cm}

{\bf 3.2. Application to concrete potentials }
\vspace{0.5cm}

Let us now investigate the system of equations (23) obtained from the
Schwinger--Dyson equation (15)
for different potentials. First of all we rewrite the potential
(13) in momentum space:
\begin{eqnarray}
V({\bf p} - {\bf q}) =
-(\frac{4}{3} \alpha_s) \frac{4 \pi}{ ({\bf{p} - \bf{q} })^2  }
-a \frac{8 \pi}{( {\bf{p} - \bf{q}})^4 } + V_0
-b (2\pi)^3 \Delta_{ {\bf q}} \delta ({\bf p} - {\bf q}) \,\,\, .
\end{eqnarray}
The last term in this potential -- the oscillator term -- should
be considered separately. Inserting it into the system (23) one gets
\begin{eqnarray}
\left\{ \begin{array} {ll}
E({\bf p}) \mbox{sin}\phi ({\bf p}) = m_0 - {\displaystyle \frac{b}{2}
\int } d {\bf q} \bigl( \Delta_{\bf q} \delta ( {\bf p}- {\bf q} ) \bigr)
\mbox{sin}\phi({\bf q}) \\ \\
E({\bf p}) \mbox{cos}\phi ({\bf p})= \vert {\bf p} \vert - {\displaystyle
\frac{b}{2} \int }
d {\bf q} \bigl( \Delta_{\bf q} \delta ( {\bf p}- {\bf q} ) \bigr)
\chi({\bf p},{\bf q} ) \mbox{cos}\phi ({\bf q}) \,\,.
\end{array}
\right.
\end{eqnarray}

Now we make use of the relations
\begin{eqnarray*}
\Delta_{\bf q} \bigl( \mbox{sin}\phi ({\bf q}) \bigr)
\vert_{{\bf p}={\bf q} } &=&
\frac{2}{p} \mbox{cos}\phi \cdot \phi^{\prime} -
\mbox{sin}\phi \cdot (\phi^{\prime})^2 +
\mbox{cos}\phi \cdot (\phi^{\prime \prime}) \,\,\, , \\
\Delta_{\bf q} \bigl( \mbox{cos}\theta ({\bf p},{\bf q}) \cdot
\mbox{sin}\phi ({\bf q}) \bigr)
\vert_{{\bf p}={\bf q} } &=& \mbox{cos}\theta \cdot \bigl(
- \frac{2}{p^2} \mbox{cos}\phi   - \frac{2}{p} \mbox{sin}\phi
 \cdot (\phi^{\prime})   \\ && -
\mbox{cos}\phi \cdot (\phi^{\prime })^2 - \mbox{sin}\phi
(\phi^{\prime \prime}) \bigr) \,\,\, ,
\end{eqnarray*}
where the notation $ p = \vert {\bf p} \vert  $ has been used.
Then, the system (26) corresponding to the Schwinger--Dyson equation for
the oscillator potential leads to the differential equation
\begin{eqnarray*}
- \frac{b}{2} ( p^2 \phi^{\prime} )^{\prime}
= p^3 \mbox{sin}\phi - m_0 p^2 \mbox{cos}\phi
+ \frac{b}{2}\mbox{sin}(2 \phi)
\end{eqnarray*}
on the function $\phi (p)$.\footnote{For $m_{0}=0$ this equation has already
been given in $[7]$.} Knowing it's solution one may in principle
calculate the constituent
quark mass $ m $ in dependence on two parameters -- the current mass
$m^0 $ and the potential parameter $b$. Indeed, let us consider relations
(16), (19), (20) for the case $ A({\bf p}) = m/ \vert {\bf p} \vert ,
B({\bf p}) = 0\,, $ so that one has
\begin{eqnarray}
\Sigma ({\bf p}) &\equiv& m \,\,, \,\,\,\,\,\,\,\,\,\,
E({\bf p}) = \sqrt{ {\bf{p}}^{2} + m^{2}} \,\,,\nonumber \\ \\
\mbox{cos} \phi({\bf p}) &=& \frac{ \vert {\bf p} \vert }
{E({\bf p})} \,\,,\,\,\,\,\,\,\,\,\,\,
\mbox{sin} \phi ({\bf p})= \frac{m}{E({\bf p})} \,\, . \nonumber
\end{eqnarray}
Here the last relation is the equation on the constituent mass $ m $.

Now we turn to the linear and Coulomb potentials in (25). In this case one
has to investigate the renormalized Schwinger--Dyson equation because the
second equation of system (23) contains an ultraviolet divergence.
We introduce into the latter a renormalization constant $ Z $,
\begin{eqnarray*}
E({\bf p})\mbox{cos}\phi({\bf p})= Z \vert {\bf p} \vert
+ \frac{1}{2} \int \frac{d {\bf{q}}}{(2\pi)^{3}} V({\bf{p}}-{\bf{q}}) \,
\chi({\bf p},{\bf q}) \, \mbox{cos}\phi({\bf q}) \,\,.
\end{eqnarray*}
and choose it as
\begin{eqnarray*}
Z=1-\frac{1}{2 \vert {\bf p} \vert} \int \frac{d {\bf{q}}}{(2\pi)^{3}}
V({\bf{p}-\bf{q}}) \, \chi({\bf p},{\bf q})  \,\,.
\end{eqnarray*}
Then the system (23) can be represented in the following manner
\begin{eqnarray}
 \left \{ \begin {array} {ll}
E({\bf p})\mbox{sin}\phi({\bf p}) = m_0 + {\displaystyle \frac{1}{2}
\int \frac{d {\bf q}}{(2 \pi)^3} }
V ( {\bf p} - {\bf q}) \mbox{sin}\phi({\bf q})  \\
E({\bf p})\mbox{cos}\phi({\bf p})
= \vert {\bf p} \vert + {\displaystyle \frac{1}{2} \int \frac{d {\bf q}}
{(2 \pi)^3} }
V ( {\bf p} - {\bf q}) \chi({\bf p},{\bf q}) ( \mbox{cos}\phi(q) -1) \,\, .
\end{array}
\right.
\end{eqnarray}
{}From here the energy $ E({\bf p}) $ may also be expressed via the function
$ \phi({\bf p}) $:
\begin{eqnarray}
E({\bf p})& = &m_{0} \mbox{sin}\phi({\bf p}) + \vert {\bf p} \vert
\mbox{cos}\phi ({\bf p}) \nonumber \\
&+& \frac{1}{2}
\int \frac{d {\bf q}}{(2\pi)^{3}} V ( {\bf p} - {\bf q}) \bigl\{
\mbox{sin}\phi({\bf p}) \mbox{sin}\phi({\bf q}) - \chi({\bf p},{\bf q})
\mbox{cos}\phi({\bf p}) (\mbox{cos}\phi({\bf q})-1) \bigr\} \,\,. \,\,\,\,\,
\,\,\,\,\,
\end{eqnarray}
The infrared singularity appearing in (28) and (29) for the Coulomb potential
may be removed if one changes the physical
observable -- the excitation energy $ \Delta E({\bf p}) = E({\bf p})
- E({\bf 0}) $.

Furthermore, from (28) one obtains the renormalized integral equation
\begin{eqnarray}
\vert {\bf p} \vert \mbox{sin}\phi({\bf p}) = m_{0} &+& \frac{1}{2}
\int \frac{d {\bf q}}{(2 \pi)^3} V ( {\bf p} - {\bf q})
\bigl\{ \mbox{cos}\phi({\bf p}) \mbox{sin} \phi({\bf q}) \nonumber \\
&& \,\,\,\,\,\,\,\,\,\,\,\,\,\,\,\,\,\,\,\,\,\,\,\,
- \chi({\bf p},{\bf q}) \mbox{sin}\phi({\bf p})( \mbox{cos}\phi({\bf q}) - 1)
\bigr\} \,\,.
\end{eqnarray}
Now, we integrate in (30) over the angular variables. Then, for the Coulomb
potential one gets
\begin{eqnarray*}
\vert {\bf p} \vert \mbox{sin}\phi({\bf p}) &=& m_{0} \mbox{cos}\phi({\bf p})
+ \frac{1}{2\pi}
\frac{4\alpha_{s}}{3} \int d \vert {\bf q} \vert \biggl\{ \biggl(
\frac{\vert {\bf q} \vert}{\vert {\bf p} \vert}
\mbox{ln} \bigg\vert \frac{\vert {\bf q} \vert -\vert {\bf p} \vert}
                          {\vert {\bf q} \vert +\vert {\bf p} \vert}
\bigg\vert \mbox{cos}\phi({\bf p})
\mbox{sin}\phi({\bf q}) \nonumber \\
&& \,\,\,\,\,\,\,\, - \biggl[ \frac{\vert {\bf q} \vert}{\vert {\bf p} \vert}
+ \frac{1}{2} \frac{{\bf q}^{2}+{\bf p}^{2}}{{\bf p}^{2}}
\mbox{ln} \bigg\vert \frac{\vert {\bf q} \vert - \vert {\bf p} \vert}
                          {\vert {\bf q} \vert + \vert {\bf p} \vert}
\bigg\vert \biggr] \mbox{sin}\phi({\bf p})
(\mbox{cos}\phi({\bf q})-1) \biggr\} \,\,.
\end{eqnarray*}
In the case of the linear potential equation (30) becomes
\begin{eqnarray*}
\vert {\bf p} \vert \mbox{sin}\phi({\bf p}) &=& m_{0} \mbox{cos}\phi({\bf p})
- \int d \vert {\bf q} \vert \biggl\{ \biggl( \frac{{\bf q}^{2}}{({\bf q}^{2}
- {\bf p}^{2})^{2}}
\mbox{cos}\phi({\bf p}) \mbox{sin}\phi({\bf q}) \nonumber \\
&& \,\,\,\,\,\,\,\,\, + \biggl[ \frac{\vert {\bf q} \vert}{\vert {\bf p}
\vert} \frac{{\bf q}^{2}+{\bf p}^{2}}{({\bf q}^{2}-{\bf p}^{2})^{2}}
+ \frac{1}{2{\bf p}^{2}} \mbox{ln} \bigg\vert
\frac{\vert {\bf q} \vert - \vert {\bf p} \vert}
     {\vert {\bf q} \vert + \vert {\bf p} \vert} \bigg\vert \biggr]
\mbox{sin}\phi({\bf p}) \mbox{cos}\phi({\bf q}) \biggr\} \,\,.
\end{eqnarray*}
The singularity in this equation can be regularized by introducing a small
cut-off $ \lambda $ around the singular point $ {\bf p}= {\bf q} $ and
taking the limit $ \lambda \rightarrow 0 $ afterwards.

\vspace{0.5cm}

{\large \bf 4. Equations on bound state masses}

\vspace{0.5cm}

{\bf 4.1. Equations for bound state vertex functions}

\vspace{0.5cm}

The Bethe--Salpeter equation (10) for the vertex functions $ \Gamma $ reads
in the case of the potential kernel (11) in momentum space
\begin{eqnarray}
\Gamma({\bf p} \vert {\cal{P}}) &=& -i \int \frac{d^{4}q}{(2\pi)^{4}}
V({\bf{p}-\bf{q}})
\rlap/{\eta} G_{1} \biggl( q+\frac{{\cal{P}}}{2} \biggr) \Gamma({\bf q} \vert
{\cal{P}}) G_{2} \biggl( q-\frac{{\cal{P}}}{2} \biggr) \rlap/{\eta} \,\,,
\end{eqnarray}
where $ G_{n} \equiv G_{\Sigma_{n}} $ is given by (17). Here the index
$ n=1,2 $ is used to distinguish between the two quarks forming the bound
state. Therefore the quantities $ \Sigma, A, B, \Lambda, S, E, \phi, \nu $
defined by (15)--(20) will now carry this index. The quantity $\cal{P}$
denotes as before the total momentum of the bound state.

For what follows we will investigate equation (31) in the rest frame,
$ \eta_{\mu} = (1,0,0,0) $:
\begin{eqnarray}
\Gamma({\bf p}) = -i \int \frac{d^{4}q}{(2\pi)^{4}} V({\bf{p}-\bf{q}})
\gamma_{0} G_{1} \biggl( q+\frac{M}{2} \biggr) \Gamma({\bf q})
G_{2} \biggl( q-\frac{M}{2} \biggr) \gamma_{0} \,\,.
\end{eqnarray}
Here $ M $ means the bound state mass. Then, according to (17) the Green's
functions $ G_{n} $ may be expressed as
\begin{eqnarray}
G_{n}(q\pm\frac{M}{2}) &=& \Biggl( \frac{\Lambda^{(n)}_{+}({\bf q})}
{q_{0}\pm M/2 -E_{n}({\bf q})+ i\epsilon} + \frac{\Lambda^{(n)}_{-}({\bf q})}
{q_{0}\pm M/2 +E_{n}({\bf q})- i\epsilon} \Biggr) \gamma_{0} \nonumber \\ \\
&=&\gamma_{0} \Biggl( \frac{\bar{\Lambda}^{(n)}_{+}({\bf q})}
{q_{0} \pm M/2-E_{n}({\bf q})+ i\epsilon} + \frac{\bar{\Lambda}^{(n)}_{-}
({\bf q})} {q_{0} \pm M/2+E_{n}({\bf q})- i\epsilon} \Biggr) \,\,, \nonumber
\end{eqnarray}
whereby $E_{n}({\bf q}), n=1,2$ are the solutions of the system of equations
(23). Integrating the Bethe--Salpeter equation (32) over $q_{0}$
we obtain the Salpeter equation
\begin{eqnarray}
\Gamma({\bf p}) = \int \frac{d{\bf{q}}}{(2\pi)^{3}} V({\bf{p}-\bf{q}})
\gamma_{0} \biggl( \frac{\Pi_{+-}({\bf q})}{E({\bf q})-M}+
\frac{\Pi_{-+}({\bf q})}{E({\bf q})+M} \biggr) \gamma_{0}
\end{eqnarray}
with $ E({\bf q})=E_{1}({\bf q})+E_{2}({\bf q}) $ and
\begin{eqnarray}
\Pi_{\pm \mp}({\bf q}) = \Lambda^{(1)}_{\pm}({\bf q}) \gamma_{0}
\Gamma \gamma_{0} \bar{\Lambda}^{(2)}_{\mp}({\bf q}) \,\, .
\end{eqnarray}

Let us now decompose $ \Gamma({\bf p}) $ with respect to it's Dirac structure,
\begin{eqnarray*}
\Gamma = \Gamma_{1}+ \gamma_{0} \Gamma_{2},
\end{eqnarray*}
where
\begin{eqnarray*}
\Gamma_{l}=\gamma^{S} \cdot \Gamma^{S}_{l}+\gamma^{P} \cdot \Gamma^{P}_{l}
+\gamma^{V}_{i}\cdot \Gamma^{V i}_{l}+ \gamma^{A}_{i} \cdot \Gamma^{Ai}_{l}
\,\,,\,\,\,\,\,\,l=1,2, \,\,\, i=1,2,3 \,\,.
\end{eqnarray*}
Here we have defined $\gamma^{S}=1,\,\gamma^{P}=\gamma_{5},\,\gamma^{V}_{i}=
\gamma_{i}, \,\gamma^{A}_{i}=\gamma_{i}\gamma_{5} $ for scalar, pseudoscalar,
vector and axial--vector bound states $ \Gamma^{S},\,\Gamma^{P},\,\Gamma^{V},
\,\Gamma^{A} $, respectively.
It is favourable to write $\Gamma$ in the form
\begin{eqnarray}
\Gamma=\sum_{I=S,P,V,A}(\Gamma^{I}_{1}+\gamma_{0}\Gamma^{I}_{2})\gamma^{I}
\,\,,
\end{eqnarray}
so that the expression $(\gamma_{0}\Gamma\gamma_{0})$ in (34) may be rewritten
as
\begin{eqnarray}
(\gamma_{0}\Gamma\gamma_{0})&=&\sum_{I=S,P,V,A}\alpha_{I}(\Gamma^{I}_{1}
                                 +\gamma_{0}\Gamma^{I}_{2})\gamma^{I}
\end{eqnarray}
with
\begin{eqnarray*}
\{\alpha_{I}, \,\, I=S,P,V,A\} = \{1,-1,-1,1\} \,\,.
\end{eqnarray*}
Then, inserting (37) into (35) one can derive from (34) the Bethe--Salpeter
equations for the vertex functions $ \Gamma^{I}_{l} $. Here we shall give only
the result of the calculation for the simple case, in which the relations
(27) hold for both quarks. Then the Green's functions (17) are given by
$ G_{n}(q)=(\rlap/{q}-m_{n}+i \epsilon)^{-1} \,,\,n=1,2 $ with $ m_{n} $ as
constituent quark mass. One obtains
\begin{eqnarray*}
\Gamma^{I}_{1}({\bf p}) &=& \int \frac{d{\bf{q}}}{(2\pi)^{3}}
V({\bf{p}-\bf{q}})
\frac{1}{E^{2}-M^{2}} \biggl\{ \biggl[(m_{1}-\alpha_{I}m_{2})
\biggl(\frac{m_{1}}{E_{1}}-\frac{m_{2}}{E_{2}} \biggr) \nonumber \\
&&\,\,\,\,\,+ (1-\alpha_{I}\beta_{I}) {\bf{q}}^{2}\biggl(\frac{1}{E_{1}}
+\frac{1}{E_{2}} \biggr)\biggr] \Gamma^{I}_{1}({\bf q})+
M \biggl(\frac{m_{1}}{E_{1}}-\alpha_{I}\frac{m_{2}}{E_{2}} \biggr)
\Gamma^{I}_{2}({\bf q}) \biggr\} \,\,, \nonumber \\  \\
\Gamma^{I}_{2}({\bf p}) &=& \int \frac{d{\bf{q}}}{(2\pi)^{3}}
V({\bf{p}-\bf{q}}) \frac{1}{E^{2}-M^{2}} \biggl\{ M
\biggl(\frac{m_{1}}{E_{1}}-\alpha_{I}\frac{m_{2}}{E_{2}} \biggr)
\Gamma^{I}_{1}({\bf q}) \nonumber \\
&& \,\,\,\,\,+ \biggl[(m_{1}-\alpha_{I}m_{2})
\biggl(\frac{m_{1}}{E_{1}}-\frac{m_{2}}{E_{2}} \biggr)
+ (1+\alpha_{I}\beta_{I}) {\bf{q}}^{2}\biggl(\frac{1}{E_{1}}+\frac{1}{E_{2}}
\biggr)\biggr] \Gamma^{I}_{2}({\bf q}) \biggr\}  \nonumber
\end{eqnarray*}
with $ E_{n} \equiv E_{n}({\bf q})=\sqrt{{\bf q}^{2}+m_{n}^{2}} \,,\,n=1,2,\,\,
E=E_{1}+E_{2} $ and
\begin{eqnarray*}
\{\beta_{I},\,\,I=S,P,V,A\}=\{-1,1,\frac{1}{3},-\frac{1}{3}\} \,\,.
\end{eqnarray*}

\vspace{0.5cm}
{\bf 4.2. Equations for bound state wave functions}
\vspace{0.5cm}

Sometimes it is favourable to work not with the vertex function $\Gamma$ but
with the Bethe--Salpeter wave function $ \Psi $. For our potential theory
(9), (11) both quantities are connected with each other in the rest frame
by the relation
\begin{eqnarray}
\Gamma({\bf p})= \int \frac{d{\bf{q}}}{(2\pi)^{3}} V({\bf{p}-\bf{q}})
\gamma_{0} \Psi({\bf q}) \gamma_{0} \,\,,
\end{eqnarray}
such that according to (34) $\Psi $ is defined as
\begin{eqnarray}
\Psi({\bf q})=\frac{\Pi_{+-}({\bf q})}{E({\bf q})-M}
            + \frac{\Pi_{-+}({\bf q})}{E({\bf q})+M}
\end{eqnarray}
with $\Pi_{\pm\mp}({\bf q})$ given by (35). Furthermore, it is more suitable
to introduce instaed of (39) a new wave function
\begin{eqnarray}
\stackrel{0}{\Psi}({\bf q})=S_{1}({\bf q})\Psi({\bf q})S_{2}({\bf q}) \,\,.
\end{eqnarray}
Then one may express the quantities $\Pi_{\pm\mp}({\bf q})$ appearing in
the definition (39) of the wave function $\Psi({\bf q})$ as follows:
\begin{eqnarray}
\Pi_{\pm\mp}({\bf q})=S_{1}^{-1}({\bf q})\stackrel{0}{\Pi}_{\pm\mp}({\bf q})
S_{2}^{-1}({\bf q})
\end{eqnarray}
with
\begin{eqnarray}
\stackrel{0}{\Pi}_{\pm\mp}({\bf q})=\stackrel{0}{\Lambda}_{\pm}
\stackrel{0}{\Gamma}({\bf q}) \stackrel{0}{\Lambda}_{\mp} \,\,,
\end{eqnarray}
where $ \stackrel{0}{\Lambda}_{\pm} $ is defined by (21) and
\begin{eqnarray}
\stackrel{0}{\Gamma}({\bf q}) = S_{1}^{-1}({\bf q}) \Gamma({\bf q})
S_{2}^{-1}({\bf q}) \,\,.
\end{eqnarray}
With the help of these relations one obtains from (38) a Salpeter
equation for the new quantities. It reads
\begin{eqnarray}
\stackrel{0}{\Gamma}({\bf p})= - \int \frac{d{\bf{q}}}{(2\pi)^{3}}
V({\bf{p}-\bf{q}}) S^{\prime}_{1}({\bf p,q}) \stackrel{0}{\Psi}({\bf q})
S^{\prime}_{2}({\bf q,p}) \,\,,
\end{eqnarray}
where
\begin{eqnarray*}
S^{\prime}_{n}({\bf p,q}) = S_{n}^{-1}({\bf p})\gamma_{0} S_{n}^{-1}({\bf q})
\,\,, \,\,\,\,\, n=1,2\,\,.
\end{eqnarray*}
Notice, that according to (40) and (41) the wave function
$\stackrel{0}{\Psi}({\bf q})$ has a representation in analogy to relation (39)
for $\Psi({\bf q})$:
\begin{eqnarray*}
\stackrel{0}{\Psi}({\bf q})=\frac{\stackrel{0}{\Pi}_{+-}({\bf q})}
{E({\bf q})-M} + \frac{\stackrel{0}{\Pi}_{-+}({\bf q})}{E({\bf q})+M} \,\,.
\end{eqnarray*}
{}From here one gets the two relations
\begin{eqnarray}
\stackrel{0}{\Lambda}_{\pm}\stackrel{0}{\Psi}({\bf q})\stackrel{0}{\Lambda}
_{\mp}= \frac{1}{E({\bf q})\mp M}\stackrel{0}{\Pi}_{\pm \mp}({\bf q}) \,\,.
\end{eqnarray}
Let us rewrite them in a form similar to Schr\"odinger equations
\begin{eqnarray}
[E({\bf q})\mp M]\stackrel{0}{\Lambda}_{\pm}\stackrel{0}{\Psi}({\bf q})
\stackrel{0}{\Lambda}_{\mp} = \stackrel{0}{\Pi}_{\pm\mp}({\bf q}) \,\,.
\end{eqnarray}

Next, we decompose the wave function $ \stackrel{0}{\Psi}({\bf q}) $ of
two--particle bound states,
\begin{eqnarray*}
\stackrel{0}{\Psi} = \stackrel{0}{\Psi}_{1} + \gamma_{0}
                     \cdot \stackrel{0}{\Psi}_{2} \,\,,
\end{eqnarray*}
and expand $\stackrel{0}{\Psi_{l}} ,\, l=1,2 $ in analogy to (36) over the
full system $\{ \gamma^{J},\,J=1,2,3\}=\{\gamma_{5},\hat{e}^{a}(p),\hat{p}\}$,
so that
\begin{eqnarray}
\stackrel{0}{\Psi} = \sum_{J=1}^{3} [\stackrel{0\,\,\,\,\,}{\Psi_{1}^{J}}
+ \gamma_{0} \cdot \stackrel{0\,\,\,\,\,}{\Psi_{2}^{J}} ] \gamma^{J} \,\,.
\end{eqnarray}
After inserting decomposition (47) into the Schr\"odinger--like equations (46)
one may derive the following systems of equations on $ \stackrel{0\,\,\,\,\,}
{\Psi_{1}^{J}} $ and $ \stackrel{0\,\,\,\,\,}{\Psi_{2}^{J}} $ for
$ J=1,2,3 $:
\begin{eqnarray}
M\stackrel{0\,\,\,\,\,}{\Psi_{2}^{J}}({\bf p}) \cdot \frac{1}{4} {\rm{tr}}
(\gamma^{K} \gamma^{J} )
&=& E({\bf p})\stackrel{0\,\,\,\,\,}{\Psi_{1}^{J}}({\bf p}) \cdot \frac{1}{4}
{\rm{tr}} (\gamma^{K} \gamma^{J} ) \nonumber \\
&-& \int \frac{d {\bf{q}}}{(2 \pi)^{3}} V({\bf{p} - \bf{q}}) T^{KL}_{12}
({\bf p,q}) \stackrel{0\,\,\,\,\,}{\Psi_{1}^{L}}({\bf q}) \,\,, \nonumber \\ \\
M\stackrel{0\,\,\,\,\,}{\Psi_{1}^{J}}({\bf p}) \cdot \frac{1}{4} {\rm{tr}}
(\gamma^{K} \gamma^{J} )
&=& E({\bf p})\stackrel{0\,\,\,\,\,}{\Psi_{2}^{J}}({\bf p}) \cdot \frac{1}{4}
{\rm{tr}} (\gamma^{K} \gamma^{J} ) \nonumber \\
&-& \int \frac{d {\bf{q}}}{(2 \pi)^{3}} V({\bf{p} - \bf{q}}) T^{KL}_{12}
({\bf p,q}) \stackrel{0\,\,\,\,\,}{\Psi_{2}^{L}}({\bf q}) \,\,, \nonumber
\end{eqnarray}
where
\begin{eqnarray*}
T^{KL}_{12} ({\bf p,q}) = \frac{1}{4} {\rm{tr}} (\gamma^{K} S^{\prime}_{1}
({\bf p,q}) \gamma^{L} S^{\prime}_{2} ({\bf q,p})) \,\,.
\end{eqnarray*}
The calculation of the traces yields
\begin{eqnarray*}
T^{KL}_{12} ({\bf p,q}) = c^{\rho_{K}}_{\bf p} c^{\rho_{L}}_{\bf q} {\rm{tr}}
(\gamma^{L} \gamma^{K} ) - s^{\rho_{K}}_{\bf p} s^{\rho_{L}}_{\bf q} {\rm{tr}}
(\gamma^{L} \gamma^{K} {\bf \hat{p} \hat{q}} )
\end{eqnarray*}
with
\begin{eqnarray*}
 \{ \rho_{K} ; K=1,2,3 \} = \{-1,-1,1\}
\end{eqnarray*}
and
\begin{eqnarray*}
c^{\pm \rho_{K}}_{\bf q} & =& c_{2}({\bf q}) c_{1}({\bf q}) \mp \rho_{K}
s_{2}({\bf q}) s_{1}({\bf q}) \,\,, \\
s^{\pm \rho_{K}}_{\bf q} & =& s_{2}({\bf q}) c_{1}({\bf q}) \mp \rho_{K}
s_{2}({\bf q}) c_{1}({\bf q}) \,\,, \\
s_{n}({\bf q})& =& \mbox{sin} \phi_{n}({\bf q}) \,\,, \,\,\,\,\,\,\,
c_{n}({\bf q}) = \mbox{cos} \phi_{n}({\bf q}) \,\,, \,\,\,\,\,\,\,n=1,2 \,\,.
\end{eqnarray*}

The systems (48) of integral equations for the bound state wave functions
$ \stackrel{0\,\,\,\,\,}{\Psi_{1}^{J}} $ and $ \stackrel{0\,\,\,\,\,}
{\Psi_{2}^{J}} $ need to be specified for every particle type.
To do this let us introduce the notations
\begin{eqnarray}
\stackrel{0\,\,\,\,\,}{\Psi_{l}^{1}}=\stackrel{0}{L}_{l} \,\,,\,\,\,\,\,
\stackrel{0\,\,\,\,\,}{\Psi_{l}^{2}}=\stackrel{0\,\,\,\,\,}{N_{l}^{a}}
\,\,,\,\,\,\,\,
\stackrel{0\,\,\,\,\,}{\Psi_{l}^{3}}=\stackrel{0}{\Sigma}_{l} \,\,,\,\,\,\,
l=1,2\,\,,
\end{eqnarray}
for pseudoscalar, vector and scalar particles, respectively. Then, for
pseudoscalar particles (48) reads
\begin{eqnarray}
M\stackrel{0}{L}_{2}({\bf{p}}) &=& E({\bf{p}})\stackrel{0}{L}_{1}({\bf{p}})
- \int \frac{d {\bf{q}}}{(2 \pi)^{3}} V({\bf{p} - \bf{q}}) (c^{-}_{\bf p}
c^{-}_{\bf q}
- \chi s^{-}_{\bf p} s^{-}_{\bf q} ) \stackrel{0}{L}_{1}({\bf{q}}) \,\,,
\nonumber \\  \\
M\stackrel{0}{L}_{1}({\bf{p}}) &=& E({\bf{p}})\stackrel{0}{L}_{2}({\bf{p}})
- \int \frac{d {\bf{q}}}{(2 \pi)^{3}} V({\bf{p} - \bf{q}}) (c^{+}_{\bf p}
c^{+}_{\bf q}
- \chi s^{+}_{\bf p} s^{+}_{\bf q} ) \stackrel{0}{L}_{2}({\bf{q}}) \,\,.
\nonumber
\end{eqnarray}
Here we have introduced as short--hand notation $s^{\pm} \equiv s^{\pm 1},\,\,
c^{\pm} \equiv c^{\pm 1}$. The quantity $ \chi \equiv \chi({\bf p},{\bf q}) $
is given by (24). For vector particles system (48) takes the form
\begin{eqnarray}
M\stackrel{0\,\,\,\,\,}{N_{2}^{a}}({\bf{p}}) &=& E({\bf{p}})
\stackrel{0\,\,\,\,\,}{N_{1}^{a}}({\bf{p}})
+ \int \frac{d {\bf{q}}}{(2 \pi)^{3}} V({\bf{p} - \bf{q}}) \nonumber \\
&&\hspace{2cm} \cdot \Big\{
(c^{-}_{\bf p} c^{-}_{\bf q} {\underline{\delta}}^{ab}
- s^{-}_{\bf p} s^{-}_{\bf q} ({\underline{\delta}}^{ab} \chi
- \eta^{a} {\underline{\eta}}^{b} )) \stackrel{0\,\,\,\,\,}{N_{1}^{b}}
({\bf{q}})
+ \eta^{a} c^{-}_{\bf p} c^{+}_{\bf q} \stackrel{0}{\Sigma}_{1}({\bf{q}})
\Bigr\} \,\,,
\nonumber \\  \\
M\stackrel{0\,\,\,\,\,}{N_{1}^{a}}({\bf{p}}) &=& E({\bf{p}})
\stackrel{0\,\,\,\,\,}{N_{2}^{a}}({\bf{p}})
+ \int \frac{d {\bf{q}}}{(2 \pi)^{3}} V({\bf{p} - \bf{q}}) \nonumber \\
&&\hspace{2cm} \cdot \Big\{
(c^{+}_{\bf p} c^{+}_{\bf q} {\underline{\delta}}^{ab}
- s^{+}_{\bf p} s^{+}_{\bf q}({\underline{\delta}}^{ab} \chi
- \eta^{a} {\underline{\eta}}^{b} )) \stackrel{0\,\,\,\,\,}{N_{2}^{b}}
({\bf{q}})
+ \eta^{a} c^{+}_{\bf p} c^{-}_{\bf q} \stackrel{0}{\Sigma}_{2}
({\bf{q}}) \Bigr\} \,\,,
\nonumber
\end{eqnarray}
where we have used the definitions
\begin{eqnarray*}
\eta^{a} \equiv \eta^{a}({\bf p},{\bf q})=\hat{q}_{i} \hat{e}^{a}_{i}({\bf p})
\,\,, \,\,\,\,\,\, {\underline{\eta}}^{a} \equiv
{\underline{ \eta}}^{a}({\bf p},{\bf q})=\hat{p}_{i} \hat{e}^{a}_{i}({\bf q})
\,\,, \,\,\,\,\,\, {\underline{\delta}}^{ab} \equiv
{\underline{ \delta}}^{ab}({\bf p},{\bf q})=\hat{e}^{a}_{i}({\bf q})
\hat{e}^{a}_{i}({\bf p}) \,\,.
\end{eqnarray*}
And for scalar particles one has
\begin{eqnarray}
M\stackrel{0}{\Sigma}_{2}({\bf{p}}) &=& E({\bf{p}})
\stackrel{0}{\Sigma}_{1}({\bf{p}}) \nonumber \\
&+& \int \frac{d {\bf{q}}}{(2 \pi)^{3}} V({\bf{p} - \bf{q}}) \Big\{
(\chi c^{+}_{\bf p} c^{+}_{\bf q} - s^{+}_{\bf p} s^{+}_{\bf q} )
\stackrel{0}{\Sigma}_{1}({\bf{q}})
+ {\underline{\eta}}^{b} c^{-}_{\bf p} c^{+}_{\bf q} \stackrel{0\,\,\,\,\,}
{N_{1}^{b}}({\bf{q}}) \Bigr\} \,\,, \nonumber \\  \\
M\stackrel{0}{\Sigma}_{1}({\bf{p}}) &=& E({\bf{p}})
\stackrel{0}{\Sigma}_{2}({\bf{p}}) \nonumber \\
&+& \int \frac{d {\bf{q}}}{(2 \pi)^{3}} V({\bf{p} - \bf{q}}) \Big\{
(\chi c^{-}_{\bf p} c^{-}_{\bf q} - s^{-}_{\bf p} s^{-}_{\bf q} ) \stackrel{0}
{\Sigma}_{2}({\bf{q}})
+ {\underline{\eta}}^{b} c^{+}_{\bf p} c^{-}_{\bf q} \stackrel{0\,\,\,\,\,}
{N_{2}^{b}}({\bf{q}})
\Bigr\} \,\,. \nonumber
\end{eqnarray}
The relations (50)--(52) for bound state wave functions have a very compact
form. In principle, they may be solved in dependence on the concrete form of
the underlying potential. Thereby, in general an exact solution would be
possible only by numerical calculations. But one can investigate also
approximate solutions for different limiting procedures as, for example,
nonrelativistic limits and the heavy quark mass limit. Let us add,
that in $[7]$ similar equations have been obtained for the oscillator
potential, and the light meson mass spectrum has been calculated numerically.

\newpage

{\large \bf 5. Meson decay constants}

\vspace{0.5cm}

In the remaining sections of this paper we want to apply the bilocal meson
model (9) with the relativistic covariant written kernel (11) to the
investigation of heavy meson properties. Therefore we have to include into
our $ QCD $--motivated model the weak interaction. We will restrict ourselves
here to the discussion of semileptonic weak processes. This allows us first
of all to determine meson decay constants.
Let us consider the quadratic part
\begin{eqnarray}
W_{eff}^{(2)} = -i \frac{N_{c}}{2} {\rm{Tr}} (G_{\Sigma} {\cal{M}})^{2}
\end{eqnarray}
of the effective action (9). First of all we expand the
bilocal field $\cal{M}$ over creation $(a^{+}_{H})$ and annihilation $(a_{H})$
operators
\begin{eqnarray}
{\cal{M}} (x,y) & = & {\cal{M}} \biggl( x-y \vert \frac{x+y}{2} \biggr)
\nonumber \\
& = & \sum_{H} \int \frac{d {\overrightarrow{\cal{P}}}}{(2 \pi)^{3/2}
\sqrt{2 \omega_{H}}}
\int \frac{d^{4}q}{(2 \pi)^{4}} e^{iq(x-y)} \Bigl\{ e^{i {\cal{P}}
\frac{x+y}{2}} a^{+}_{H}({\bf{q}} \vert {\cal{P}})
\Gamma_{H}({\bf{q}} \vert {\cal{P}}) \\
&& \hspace{6.0cm} + e^{-i {\cal{P}} \frac{x+y}{2}}
a_{H}({\bf{q}} \vert {\cal{P}})
\bar{\Gamma}_{H}({\bf{q}} \vert {\cal{P}}) \Bigr\} \,\,. \nonumber
\end{eqnarray}
Here the sum runs over the set of quantum numbers $H$ of hadrons contributing
in the bilocal fields $ {\cal{M}} (x,y) $ . The bound state has the total
momentum $ {\cal P} = \{\omega_{H},{\overrightarrow{\cal P}} \} \,,$ the
energy $ \omega_{H}(\overrightarrow{\cal{P}}) =
\sqrt{{\overrightarrow{\cal{P}}}^{2}+ M_{H}^{2}} $
and the mass $ M_{H} $. The bound state vertex functions $ \Gamma_{H} $ and
$ \bar{\Gamma}_{H} $ satisfy the Bethe--Salpeter equation (10) with kernel
(11).

Furthermore, we have to include the weak interactions into the effective
action (53). The effective Lagrangian of semileptonic weak interaction has
the form
\begin{eqnarray}
{\cal{L}}_{semi} = \frac{G_{F}}{\sqrt{2}} \{V_{ij}(\bar{Q}_{i} O_{\mu} q_{j})
l_{\mu}  + h.c. \}
\end{eqnarray}
with the leptonic current
\begin{eqnarray*}
l_{\mu} \equiv \bar{l} O_{\mu}\nu_{l} \,\,,\,\,\,\, l=e,\mu,\tau \,\,,
\,\,\,\,\,\,\, O_{\mu}=\gamma_{\mu}(1+\gamma_{5}) \,\,,
\end{eqnarray*}
the elements $ V_{ij} $ of
the Kobayashi--Maskawa matrix and the Fermi constant $ G_{F}= 10^{-5}/m_{p}
^{2} $. $ Q $ denotes the column of $ (u,c,t) $ quarks and $ q $ -- the column
of $ (d,s,b) $ quarks. The lagrangian (55) can be incorporated into the
bilocal action (53) by substituting
\begin{eqnarray}
{\cal{M}}(x,y) \rightarrow {\cal{M}}(x,y) + \hat{L}(x,y) \,\,,
\end{eqnarray}
i.e., by adding to the bilocal field $ {\cal{M}}(x,y) $ the local
weak leptonic current
\begin{eqnarray}
\hat{L}_{ij}(x,y) = \frac{G_{F}}{\sqrt{2}} \delta^{4}(x-y) V _{ij} \hat{l}
e^{i{\cal{P}}_{L} \frac{x+y}{2}} \,\,,
\end{eqnarray}
where $ \hat{l} \equiv O_{\mu} l_{\mu} $ and $ {\cal{P}}_{L} $ being the
momentum of the leptonic pair.
Then, the terms of interest standing in (53) after the substitution (56) and
corresponding to semileptonic weak interaction are
\begin{eqnarray*}
W_{semi}^{(2)} &=& -i N_{c} {\rm{Tr}} (G {\cal{M}} G \hat{L} ) \\ &\equiv &
-i N_{c} \int dxdydzdt \, {\rm{tr}} \Bigl[ G_{i^{\prime}}(t-x)
{\cal{M}}_{i^{\prime}k}(x,y) G_{j^{\prime}} (y-z) \hat{L}_{j^{\prime}k}(z,t)
\Bigr] \,\,.
\end{eqnarray*}
Here the trace runs over Dirac and flavour indices.

Now we are able to derive a formula for pseudoscalar meson decay constants.
The matrix element for a decay of a meson $ H_{ij} \sim (q_{i} \bar{q}_{j}) $
into a leptonic pair reads
\begin{eqnarray}
<l \nu \vert W^{(2)}_{semi} \vert H_{ij} > &=& -i N_{c} (2 \pi )^{4} i
\delta^{(4)}({\cal{P}}_{H} -{\cal{P}}_{L} ) \frac{G_{F}}{\sqrt{2}}
<l \nu \vert l_{\mu} \vert 0 >  \nonumber \\
&& \cdot \int \frac{d^{4}q}{(2 \pi)^{4}} {\rm{tr}}_{\gamma} \biggl\{ O_{\mu}
G_{i} \biggl( q- \frac{{\cal{P}}_{H}}{2} \biggr)
\bar{\Gamma} ({\bf q} \vert {\cal{P}}_{H}) G_{j} \biggl
(q+ \frac{{\cal{P}}_{H}}{2} \biggr) \biggr\} \,\,.\,\,\,\,\,
\end{eqnarray}
Using the relations (33) for Green's functions and the definition of a bound
state wave function (from (32) and (38))
\begin{eqnarray*}
i \int \frac{dq_{0}}{2 \pi} G_{i}(q -\frac{ {\cal{P}}}{2})
\bar{\Gamma}({\bf q} \vert {\cal{P}}) G_{j}(q+ \frac {{\cal{P}}}{2})
\equiv \bar{\Psi}_{H_{ij}}({\bf q} \vert {\cal{P}}) \,\,,
\end{eqnarray*}
we can write
\begin{eqnarray}
&&\int \frac{d^{4}q}{(2 \pi)^{4}} {\rm{tr}}_{\gamma} \biggl\{ O_{\mu}
G_{i} \biggl( q- \frac{{\cal{P}}_{H}}{2} \biggr)
\bar{\Gamma}({\bf q} \vert {\cal{P}}_{H}) G_{j} \biggl(
q+ \frac{{\cal{P}}_{H}}{2} \biggr) \biggr\} \nonumber \\
&&= -i \int \frac{d {\bf{q}}}{(2 \pi)^{3}} {\rm{tr}}_{\gamma}
[ O_{\mu} {\bar{\Psi}}_{H_{ij}}({\bf q} \vert {\cal{P}}_{H}) ] \,\,,
\end{eqnarray}
whereby $ {\bar{\Psi}}({\bf q} \vert {\cal{P}}_{H}) \equiv
\Psi({\bf q} \vert -{\cal{P}}_{H}) $ .
In analogy to (40) we introduce now the "dressed" wave function $ \stackrel{0}
{\Psi} $ and expand it in correspondence with (47) over the Lorentz matrices.
Because we are interested in pseudoscalar mesons only, we take $ J=1 $ and
have then in the moving frame (using notation (49))
\begin{eqnarray}
\stackrel{0}{\Psi}({\bf q} \vert -{\cal{P}}_{H}) =
\biggl( \stackrel{0}{L}_{1}({\bf q} \vert -{\cal{P}}_{H}) - \rlap/{\eta}
\cdot \stackrel{0}{L}_{2}({\bf q} \vert -{\cal{P}}_{H}) \biggr) \gamma^{5}\,\,.
\end{eqnarray}
Remember, that $ \eta^{\mu} = {\cal{P}}^{\mu}_{H} / \sqrt{{\cal{P}}^{2}_{H}} $
with $ {\cal{P}}^{0}_{H} = M_{H} $.
Inserting (60) into (59) one gets after the calculation of the trace
\begin{eqnarray*}
i4 \eta_{\mu} \int \frac{d {\bf{q}}}{(2 \pi)^{3}}
\Bigl( \mbox{cos} \nu_{i}({\bf{q}})\cdot \mbox{cos} \nu_{j}({\bf{q}})
- \mbox{sin} \nu_{i}({\bf{q}}) \cdot \mbox{sin} \nu_{j}({\bf{q}}) \Bigr)
(\stackrel{0}{L}_{2})_{H_{ij}}({\bf{q}} \vert {\cal{P}}_{H})  \,\,,
\end{eqnarray*}
where $ \nu({\bf{q}}) $ is defined in (20).
Then, this expression is inserted into (57). Comparing the result with the
general formula
\begin{eqnarray*}
<l \nu \vert W^{(2)}_{semi} \vert H_{ij} > = (2 \pi )^{4} i
\delta^{4}({\cal{P}}_{H} -{\cal{P}}_{L} ) \frac{G_{F}}{\sqrt{2}} i
F_{H_{ij}} {\cal{P}}^{\mu}_{H} <\omega \vert l_{\mu} \vert 0 >  \,\,
\end{eqnarray*}
one obtains for the decay constant of a pseudoscalar meson $ H_{ij} $ at rest
\begin{eqnarray}
F_{H_{ij}} = \frac{4N_{c}}{M_{H}} \int \frac{d {\bf{q}}}{(2 \pi)^{3}}
(\stackrel{0}{L}_{2})_{H_{ij}}({\bf{q}}) \mbox{cos} \bigl( \nu_{i}({\bf{q}})
+\nu_{j}({\bf{q}}) \bigr)  \,\,.
\end{eqnarray}
Notice, that in the rest frame
the functions $ \stackrel{0}{L}_{1} $ and $ \stackrel{0}{L}_{2} $ satisfy the
system of equations (50).
So we have for $B$ and $D$ mesons the relations
\begin{eqnarray*}
F_{D} &=& \frac{4N_{c}}{M_{D}} \int \frac{d {\bf{q}}}{(2 \pi)^{3}}
\bigl( \stackrel{0}{L}_{2} \bigr)_{D} \mbox{cos}(\nu_{c}+\nu_{u})  \,\,;\\
F_{D_{s}} &=& \frac{4N_{c}}{M_{D_{s}}} \int \frac{d {\bf{q}}}{(2 \pi)^{3}}
\bigl( \stackrel{0}{L}_{2} \bigr)_{D_{s}} \mbox{cos}(\nu_{c}+\nu_{d})
\,\,; \\
F_{B_{u}} &=& \frac{4N_{c}}{M_{B_{u}}} \int \frac{d {\bf{q}}}{(2 \pi)^{3}}
\bigl( \stackrel{0}{L}_{2} \bigr)_{B_{u}} \mbox{cos}(\nu_{b}+\nu_{u})
\,\,; \\
F_{B_{c}} &=& \frac{4N_{c}}{M_{B_{c}}} \int \frac{d {\bf{q}}}{(2 \pi)^{3}}
\bigl( \stackrel{0}{L}_{2} \bigr)_{B_{c}} \mbox{cos}(\nu_{b}+\nu_{c}) \,\,.
\end{eqnarray*}

\vspace{0.5cm}

{\large \bf 6. The relation between bilocal field approach and
heavy quark effective theory}

\vspace{0.5cm}

For the description of the properties of heavy quarkonia one can employ the
heavy quark effective theory, which has been developed recently $[8-10]$.
Let us here consider mesons $ (Q\bar{q}) $ consisting of a heavy quark $ Q $
and a light antiquark $ \bar{q} $. Now we make use of the fact that the
quarkonium velocity $ v^{\mu} $ is determined more or less by the velocity
$ v^{\mu}_{Q} $  of its heavy quark constituent $ Q $. Then, the momentum of a
bound state of velocity $ v^{\mu} $ is given by
\begin{eqnarray*}
{\cal{P}}^{\mu} = M v^{\mu} \,\,,
\end{eqnarray*}
where $ M $ is the bound state mass. The latter is approximately equal to
the heavy quark mass $ m_{Q} $, i.e.
\begin{eqnarray*}
M \approx m_{Q} \,\,.
\end{eqnarray*}
Then one can write for the heavy quark momentum
\begin{eqnarray}
p^{\mu}_{Q} = m_{Q} v^{\mu}_{Q} = {\cal{P}}^{\mu} - p_{q}^{\mu} =
m_{Q}v^{\mu} + k^{\mu}
\end{eqnarray}
with
\begin{eqnarray}
k^{\mu} = (M-m_{Q}) v^{\mu} - p_{q}^{\mu} \,\,,
\end{eqnarray}
where $ p_{q} $ is the light antiquark momentum. From (61) it follows that
the heavy quark velocity is given by
\begin{eqnarray}
v^{\mu}_{Q} = v^{\mu} + \frac{1}{m_{Q}} k^{\mu} \,\,.
\end{eqnarray}
Therefore, in the limit $ m_{Q} \rightarrow \infty $ one has $ v^{\mu}_{Q}
\rightarrow v^{\mu} $. In (62) and (63) the light antiquark momentum
$ p_{q}^{\mu} $ is small as compared with $ {\cal{P}}^{\mu} $, and it has
only little influence on the direction of the heavy quark motion. As result
one obtains that a meson bound state containing a heavy quark can be
considered by a theory describing heavy quark motion, for which the $ QCD $
corrections are small. Therefore the bound state is formed by a
colour Coulomb field.

Let us now assume that the bilocal field $ {\cal{M}}(x,y) $ contains a heavy
quark ($ b $ or $ c $) and a light antiquark ($ \bar{u} $, $ \bar{d} $ or
$ \bar{s} $). Now, the
integral kernel $ {\cal{K}}^{\eta}(x-y) $, eq. (11), is defined with the
help of the vector $ \eta_{\mu} $, which has been introduced for two reasons.
Firstly, it defines the transversality of the interaction given effectively
by the phenomenological potential $ V(z^{\perp}) $. And secondly, the vector
$ \eta_{\mu} $ determines the motion of the interaction potential together
with the motion of the $ (Q \bar{q}) $ bound state. In the limit of the heavy
quark effective theory one has
\begin{eqnarray*}
\eta_{\mu} = \frac{{\cal{P}}_{\mu}}{\sqrt{{\cal{P}}^{2}}} \rightarrow v_{\mu}
\,\,,
\end{eqnarray*}
where $ v^{2}=1 $. Therefore the interaction kernel (11) takes in this limit
the form
\begin{eqnarray*}
{\cal{K}}^{v}(x-y)=\rlap/{v} V(z^{\perp}) \delta^{(4)}(v \cdot z) \otimes
                   \rlap/{v} \,\,,
\end{eqnarray*}
where $ z^{\parallel}_{\mu} = v_{\mu} (v \cdot z) $. We conclude that for
considering the heavy quark effective theory one has to investigate the case
of moving bound states in our bilocal field theory for $m_{Q} \rightarrow
\infty$. Concering the problems that have been discussed in this paper so far
it was sufficient to consider the bound states at rest. Let us note the main
modifications that appear if one
works in an arbitrary reference frame. Instead of the 3--momentum ${\bf{q}}$
one has now $q^{\perp}$, and $q^{0}$ is substituted by $q^{\parallel}$. The
product $q_{i}\gamma_{i}$ changes to $-\rlap/{q}^{\perp}=-q^{\perp}_{\mu}
\gamma^{\mu}$. Instead of $\vert {\bf{q}} \vert$ one has $q^{\perp}\equiv
\sqrt{(q^{\perp})^{2}} $, so that  $\hat{q}^{\perp}=\hat{q}^{\perp}_{\mu}
\gamma^{\mu}, \hat{q}^{\perp}_{\mu}= q^{\perp}_{\mu}/q^{\perp}$. Then, for
instance, eqs. (16) and (22) for $\Sigma$ read
\begin{eqnarray*}
\Sigma(q^{\perp})=A(q^{\perp})q^{\perp} + B(q^{\perp}) \rlap/{q}=
\rlap/{q}^{\perp} + E(q^{\perp})S^{-2}(q^{\perp}) \,\,.
\end{eqnarray*}

For illustration
let us now consider the Schwinger--Dyson equation in the heavy mass limit.
Thereby, we will restrict ourselves to the investigation of the case (27), in
which $ \Sigma(p^{\perp}_{Q}) \equiv m_{Q} $. Then, for $ S^{\pm 2}
(q^{\perp})$ one obtains according to (20)
\begin{eqnarray*}
 S^{\pm 2}(q^{\perp}) = \frac{m_{Q}}{E(q^{\perp})} \pm \hat{q}^{\perp}
                       \frac{\vert q^{\perp} \vert}{E(q^{\perp})} \,\,.
\end{eqnarray*}
The Schwinger--Dyson equation (15) for a heavy quark reads
\begin{eqnarray}
m_{Q} = m^{0} - i {\rm{tr}} \int \frac{d^{4}q}{(2\pi)^{4}} V(q^{\perp})
     \rlap/{v} G_{m_{Q}}(p_{Q} - q) \rlap/{v}  \,\,,
\end{eqnarray}
where the trace runs over colour and Dirac indices. In the heavy mass limit
we take inside the loop integral of (65) for the quark propagator the
expression
\begin{eqnarray}
G_{m_{Q}}(p_{Q} - q) = \frac{m_{Q}(1+\rlap/{v}) - \rlap/{q}}{2m_{Q}v(k-q)
+ q^{2} + i\epsilon}  \,\,.
\end{eqnarray}
Here we have used equation (62) and from (64) the relation
$ k^{\mu} / m_{Q} \ll v^{\mu}_{Q} \approx v^{\mu} $. Inserting (66) into (65)
one gets
\begin{eqnarray*}
m_{Q} & = & m^{0} - i 4N_{c} \int \frac{d^{4}q}{(2\pi)^{4}} V(q^{\perp})
        \frac{m_{Q}}{2m_{Q}v(k-q) + q^{2} + i\epsilon}     \nonumber \\
     & = & m^{0} + 2 N_{c} \int \frac{d^{3}q^{\perp}}{(2\pi)^{3}} V(q^{\perp})
        \frac{m_{Q}}{\sqrt{m_{Q}^{2}v^{2} - 2m_{Q}vk + (q^{\perp})^{2}}} \,\,.
\end{eqnarray*}

\vspace{0.5cm}

{\large \bf 7. Semileptonic decays of heavy quarkonia in the limit of
heavy quark effective theory}

\vspace{0.5cm}

Let us now consider semileptonic decays of heavy mesons. Therefore we have
to investigate the cubic part of the effective action (9), thereby
substituting one of the bilocal fields $ {\cal{M}} $ by the local leptonic
current $\hat{L}$ from (57):
\begin{eqnarray}
W^{(3)}_{semi} &=& i \frac{N_{c}}{3} {\rm{Tr}} [(G_{\Sigma} M)^{2}
                                             (G_{\Sigma} \hat{L})] \nonumber \\
& \equiv & \frac{iN_{c}}{3} \int \frac{d^{4}x_{1}}{(2\pi)^{4}} \cdots
    \int \frac{d^{4}x_{6}}{(2\pi)^{4}} \, {\rm{tr}} [G_{\Sigma}(x_{1}-x_{2})
   {\cal{M}}(x_{2},x_{3}) G_{\Sigma}(x_{3}-x_{4}) \nonumber  \\
&& \hspace{5.5cm}
\hat{L}(x_{4},x_{5})  G_{\Sigma}(x_{5}-x_{6}) {\cal{M}}(x_{6},x_{1})] \,\,.
\end{eqnarray}
Here the arguments in the integrand have been introduced according to
fig.1. After rewriting (67) in momentum space and using the decomposition (54)
for $ {\cal{M}}(x,y) $ one obtains for the matrix element describing a
semileptonic decay of a B--meson $H_{ib}$ with a meson
$H^{\prime}_{ji}$ in the final state
\begin{eqnarray}
< (l\nu_{l}) H^{\prime}_{ji} \vert W^{(3)}_{semi} \vert H_{ib} >
&=& -i \frac{N_{c}}{3} (2\pi)^{4} \delta({\cal{P}}-{\cal{P}}^{\prime}-
{\cal{P}}_{L}) \frac{1}{2 (2\pi)^{3} \sqrt{\omega \omega^{\prime}}} \nonumber
\\ && \cdot <l \nu \vert l_{\mu} \vert 0> \cdot \frac{G}{\sqrt{2}} V_{jb}
\, I_{bji}^{\mu}({\cal{P}},{\cal{P}}^{\prime})
\end{eqnarray}
with
\begin{eqnarray}
I_{bji}^{\mu}({\cal{P}},{\cal{P}}^{\prime}) &=& {\rm{tr}}
\int \frac{d^{4}k}{(2\pi)^{4}} G_{b}(k-{\cal{P}}) O^{\mu}
     G_{j}(k-{\cal{P}}^{\prime}) \nonumber \\
&& \hspace{2.5cm} \cdot \Gamma_{H^{\prime}}
              ({\bf{k}}-\frac{{{\bf\cal{P}}}^{\prime}}{2} \vert {\cal{P}})
     G_{i}(k)\bar{\Gamma}_{H}
              ({\bf{k}}-\frac{{\bf{\cal{P}}}}{2} \vert {\cal{P}})  \,\,,
\end{eqnarray}
$O^{\mu}=\gamma^{\mu}(1+\gamma_{5})$ and the latain indices $b,\,j=c,u$ and
$i=u,d,s,c$ indicating the quark content of the quantities. The momenta in the
integral $I_{bji}^{\mu}$ have been
introduced according to fig.2. In (69) $ G_{b}, G_{j} $ and
$ G_{i} $ are the Green's functions of the quarks.
For example, for $j=c$ and $i=u,d,s,c$ the matrix element (68) describes
the decays $ B^{-}_{u} \rightarrow D^{0}(l\nu_{l}) $,
$ B^{0}_{d} \rightarrow D^{+}(l\nu_{l}) $,
$ B^{0}_{s} \rightarrow D^{+}_{s}(l\nu_{l}) $,
and $ B^{-}_{c} \rightarrow (c \bar{c})(l\nu_{l}) $, respectively.

For definiteness let us investigate the matrix element (68) for a semileptonic
B--decay into a D--meson ($j=c$). The calculation will be done as
follows. Knowing from sect.6 the relation between bilocal field approach and
heavy quark effective theory we will work at first in the rest frame. Only at
the end we shall turn to the moving frame by substituting $\gamma_{0}$ by
$ \rlap/{\eta} $ what corresponds for $ \rlap/{\eta}=\rlap/{v} $ to the heavy
quark mass limit.

Proceeding in this way let us start with rewriting the integral (69).
At first we insert the expressions (33) for the Green's functions into the
latter:
\begin{eqnarray}
I_{bci}^{\mu}(M,M^{\prime}) &=& {\rm{tr}} \int \frac{d^{4}k}{(2\pi)^{4}} \,\,
\Biggl( \frac {\Lambda_{+}^{(b)}({\bf{k}})}{k_{0}-k_{1}^{(b)}+i\epsilon} +
       \frac {\Lambda_{-}^{(b)}({\bf{k}})}{k_{0}-k_{2}^{(b)}-i\epsilon} \Biggr)
\gamma_{0} O^{\mu} \nonumber \\ && \hspace{2cm}
\cdot \Biggl( \frac {\Lambda_{+}^{(c)}({\bf{k}})}{k_{0}-k_{1}^{(c)}+i\epsilon}
     + \frac {\Lambda_{-}^{(c)}({\bf{k}})}{k_{0}-k_{2}^{(c)}-i\epsilon} \Biggr)
\gamma_{0} \Gamma_{H}({\bf{k}}) \gamma_{0}  \nonumber  \\
&& \hspace{2cm}
\cdot \Biggl( \frac {\bar{\Lambda}_{+}^{(i)}({\bf{k}})}{k_{0}-k_{1}^{(i)}+
                                                                    i\epsilon}
 \frac {\bar{\Lambda}_{-}^{(i)}({\bf{k}})}{k_{0}-k_{2}^{(i)}-i\epsilon} \Biggr)
\bar{\Gamma}_{H^{\prime}} ({\bf{k}}) \,\,,
\end{eqnarray}
with
\begin{eqnarray*}
k_{1/2}^{(b)} &=& M \pm E_{b}({\bf{k}})  \,\,,\\
k_{1/2}^{(c)} &=& M^{\prime} \pm E_{c}({\bf{k}})  \,\,,\\
k_{1/2}^{(i)} &=& \pm E_{i}({\bf{k}}) \,\,.
\end{eqnarray*}
Now we represent the integrand of $I_{bci}$ as a sum of eight terms the
numerators of which have the form
\begin{eqnarray*}
&&{\rm{tr}} [\Lambda_{\pm}^{(b)}({\bf{k}}) \gamma_{0} O^{\mu}
           \Lambda_{\pm}^{(c)}({\bf{k}})
           \gamma_{0} \Gamma_{H}({\bf{k}}) \gamma_{0}
           \bar{\Lambda}_{\pm}^{(i)}({\bf{k}})
           \bar{\Gamma}_{H^{\prime}} ({\bf{k}})] \\
&&={\rm{tr}} [S^{-1}_{b}({\bf{k}}) O^{\mu} S^{-1}_{c}({\bf{k}})
           (\pm \stackrel{0}{\Lambda}_{\pm})
           \stackrel{0}{\Gamma}_{H}({\bf{k}})
           (\pm \stackrel{0}{\Lambda}_{\pm})
           \stackrel{0}{\bar{\Gamma}}_{H^{\prime}}({\bf{k}})
           (\pm \stackrel{0}{\Lambda}_{\pm})] \,\,.
\end{eqnarray*}
Here the last line is obtained by using (18) and (43), whereby $\stackrel{0}
{\Gamma}$ fulfils the Bethe--Salpeter equation (44). Because of the relations
\begin{eqnarray*}
\stackrel{0}{\Lambda}_{\pm} \stackrel{0}{\Gamma}_{H}
\stackrel{0}{\Lambda}_{\pm}=0 \,\,, \,\,\,\,\,
\stackrel{0}{\Lambda}_{\pm} \stackrel{0}{\Gamma}_{H}
\stackrel{0}{\Lambda}_{\mp}= \stackrel{0 \hspace{0.5cm}}{{\Pi}^{H}
                                                                  _{\pm\mp}}
\end{eqnarray*}
(cf. (42)) only two numerators remain. Then (70) is given by
\begin{eqnarray*}
I_{bci}^{\mu}(M,M^{\prime}) &=& {\rm{tr}} \int \frac{d^{4}k}{(2\pi)^{4}}
                  S^{-1}_{b}({\bf{k}}) O^{\mu} S^{-1}_{c}({\bf{k}}) \\
&& \hspace{2cm} \cdot \Biggl( \frac {\stackrel{0\hspace{0.5cm}}
                             {{\Pi}_{-+}^{H^{\prime}}}({\bf{k}})
               \stackrel{0}{\bar{\Gamma}}_{H}({\bf{k}})
               \stackrel{0}{\Lambda}_{-}}
              {(k_{0}-k_{2}^{(b)}-i \epsilon) (k_{0}-k_{1}^{(c)}+i \epsilon)
               (k_{0}-k_{2}^{(i)}-i\epsilon)} \\
&& \hspace{2.5cm} - \frac {\stackrel{0\hspace{0.5cm}}
                       {{\Pi}_{+-}^{H^{\prime}}}({\bf{k}})
               \stackrel{0}{\bar{\Gamma}}_{H}({\bf{k}})
               \stackrel{0}{\Lambda}_{+}}
              {(k_{0}-k_{1}^{(b)}+i \epsilon) (k_{0}-k_{2}^{(c)}-i \epsilon)
               (k_{0}-k_{1}^{(i)}+i\epsilon)} \Biggr) \,\,.
\end{eqnarray*}
Now we can perform the $k_{0}$ integration. The result reads
\begin{eqnarray}
I_{bci}^{\mu}(M,M^{\prime}) &=& i{\rm{tr}} \int \frac{d {\bf{k}}}{(2\pi)^{3}}
              S^{-1}_{b}({\bf{k}}) O^{\mu} S^{-1}_{c}({\bf{k}}) \nonumber \\
&& \hspace{2cm} \cdot
\Biggl( \frac {\stackrel{0\hspace{0.5cm}}{{\Pi}_{-+}^{H^{\prime}}}
({\bf{k}}) \stackrel{0}{\bar{\Gamma}}_{H}({\bf{k}})
               \stackrel{0}{\Lambda}_{-}}
     {[E_{c}+E_{b}-(M-M^{\prime})](E_{c}+E_{i}+M^{\prime}) } \nonumber \\
&& \hspace{2cm}  - \frac {\stackrel{0\hspace{0.5cm}}
                         {{\Pi}_{+-}^{H^{\prime}}}({\bf{k}})
               \stackrel{0}{\bar{\Gamma}}_{H}({\bf{k}})
               \stackrel{0}{\Lambda}_{+}}
     {[E_{c}+E_{b}+(M-M^{\prime})](E_{c}+E_{i}-M^{\prime})} \Biggr) \,\,.
\end{eqnarray}
Here the expression in the brackets can be rewritten according to (45) as
follows
\begin{eqnarray*}
&&\frac {\stackrel{0}{\Lambda}_{-} \stackrel{0}{\Psi}_{H^{\prime}}
       ({\bf{k}}) \stackrel{0}{\Lambda}_{+}
       \stackrel{0}{\bar{\Gamma}}_{H}({\bf{k}})
               \stackrel{0}{\Lambda}_{-}}
     {E_{c}+E_{i}+M^{\prime}}
- \frac {\stackrel{0}{\Lambda}_{+} \stackrel{0}{\Psi}_{H^{\prime}}
       ({\bf{k}}) \stackrel{0}{\Lambda}_{-}
       \stackrel{0}{\bar{\Gamma}}_{H}({\bf{k}})
               \stackrel{0}{\Lambda}_{+}}
     {E_{c}+E_{i}-M^{\prime}} \\
&&= \frac {\stackrel{0}{\Lambda}_{-} \stackrel{0} {\Psi}_{H^{\prime}}
 ({\bf{k}}) \stackrel{0\hspace{0.5cm}}{{\Pi}^{H}_{+-}} ({\bf{k}})}
        {E_{c}+E_{i}+M^{\prime}}
- \frac {\stackrel{0}{\Lambda}_{+} \stackrel{0}{\Psi}_{H^{\prime}}
    ({\bf{k}}) \stackrel{0\hspace{0.5cm}}{{\Pi}^{H}_{-+}}({\bf{k}})}
        {E_{c}+E_{i}-M^{\prime}} \,\,.
\end{eqnarray*}
Using (45) again and inserting the result in (71) one obtains
\begin{eqnarray}
I_{bci}^{\mu}(M,M^{\prime}) &=& i{\rm{tr}} \int \frac{d {\bf{k}}}{(2\pi)^{3}}
         S^{-1}_{b}({\bf{k}}) O_{\mu} S^{-1}_{c}({\bf{k}})
\Bigl( \stackrel{0}{\Lambda}_{-} \stackrel{0}{\Psi}_{H^{\prime}}
           ({\bf{k}}) \stackrel{0}{\Lambda}_{+}
        \stackrel{0}{\bar{\Psi}}_{H} ({\bf{k}})
                      \stackrel{0}{\Lambda}_{-}  \nonumber \\ && \hspace{5cm}
      - \stackrel{0}{\Lambda}_{+} \stackrel{0}{\Psi}_{H^{\prime}}
           ({\bf{k}}) \stackrel{0}{\Lambda}_{-}
        \stackrel{0}{\bar{\Psi}}_{H}({\bf{k}})
                      \stackrel{0}{\Lambda}_{+} \Bigr)  \,\,. \,\,\,\,\,
\end{eqnarray}

In the following we will further rewrite the integral (72) in the heavy
quark limit. To do this we have first of all to turn to the moving frame.
We introduce the momenta as shown in fig.2. In the moving frame we
consider the decomposition $ k_{\mu}=k_{\mu}^{\parallel}+k_{\mu}^{\perp} $
of the momentum $ k_{\mu}=(k_{0},{\bf{k}}) $ with respect to the momentum
$ {\cal{P}} $ of the initial boson. Furthermore, $ \gamma_{0} $ has to be
replaced by $ \rlap/{\eta}=\rlap/{v} $ for the incoming $b$ quark and by
$ \rlap/{\eta}^{\prime}=\rlap/{v}^{\prime} $ for the outgoing $c$ quark.
Therefore in the moving frame the integral (72) takes the form
\begin{eqnarray}
I_{bci}^{\mu}(v,v^{\prime})&=& - \frac{i}{2} {\rm{tr}} \int
                                   \frac{d^{3}k^{\perp}} {(2\pi)^{3}}
                  S^{-1}_{b}(k^{\perp}) O^{\mu} S^{-1}_{c}(k^{\perp})
\nonumber\\
&&\hspace{2.5cm} \cdot (1+\rlap/{v}^{\prime})
\stackrel{0}{\Psi}_{H^{\prime}}(k^{\perp})
\stackrel{0}{\bar{\Psi}}_{H} (k^{\perp}) (1+\rlap/{v})   \,\,.
\end{eqnarray}
In this relation we have also made use of the fact that for heavy quarks
the change to the moving frame leads immediately to the heavy quark effective
theory. So for heavy quark fields $Q$ with momentum ${\cal{P}}^{\mu}=m v^{\mu}
+ k^{\mu}, k^{\mu} \ll M $ one has
\begin{eqnarray*}
(1-\rlap/{v})Q({\cal{P}}) \approx \frac{\rlap/{k}}{M} Q({\cal{P}})
\approx 0   \,\,.
\end{eqnarray*}
Therefore the projector on the antipartilce vanishes: $\stackrel{0
\hspace{0.5cm}}{\Lambda_{-}^{(Q)}} \rightarrow 0 $. This means that for
heavy quarks the only relevant contribution comes from $\stackrel{0
\hspace{0.5cm}}{\Lambda_{+}^{(Q)}} \rightarrow \Lambda_{+}^{(Q)} =
(1+ \rlap/{v})/2 $. Furthermore, we have used in (73) for the light quark $q$
in the heavy quark limit
$\stackrel{0 \hspace{0.5cm}}{\Lambda_{\pm}^{(q)}} \rightarrow 1/2 $.

For definiteness let us now consider the semileptonic decay of a pseudoscalar
$B$ meson into a heavy pseudoscalar meson of the type $ (c \bar{i}), \,\,
i=u,d,s,c $. In this
case we can use the decomposition (47) of the meson wave function which for
moving fields takes the form
\begin{eqnarray*}
\stackrel{0}{\Psi}_{H^{\prime}}(k^{\perp} \vert {\cal{P}}^{\prime}) &=&
\bigl (L_{1}^{H^{\prime}} +
\rlap/{v}^{\prime} L_{2}^{H^{\prime}} \bigr)
(k^{\perp} \vert {\cal{P}}^{\prime}) \gamma_{5} \,\,, \\
\stackrel{0}{\bar{\Psi}}_{H}(k^{\perp} \vert {\cal{P}}) &=& \gamma_{5}
\bigl ( \bar{L}_{1}^{H} +
\rlap/{v} \bar{L}_{2}^{H} \bigr)
(k^{\perp} \vert {\cal{P}}) \,\,.
\end{eqnarray*}
Inserting these decompositions into (73) one gets
\begin{eqnarray}
I_{bci}^{(PS)\mu}(v,v^{\prime})&=& - \frac{i}{2} {\rm{tr}} \int
                                      \frac{d^{3}k^{\perp}} {(2\pi)^{3}}
          S^{-1}_{b}(k^{\perp}) O^{\mu} S^{-1}_{c}(k^{\perp}) \nonumber  \\
&&\hspace{2.5cm} \cdot (1+\rlap/{v}^{\prime}+\rlap/{v}
+\rlap/{v}^{\prime}\rlap/{v}) W(k^{\perp} \vert v,v^{\prime}) \,\,, \\
\nonumber \\
W(k^{\perp} \vert v,v^{\prime}) &=&
L_{1}^{H^{\prime}}(k^{\perp} \vert {\cal{P}}^{\prime})
\bar{L}_{1}^{H}(k^{\perp} \vert {\cal{P}}) +
L_{1}^{H^{\prime}}(k^{\perp} \vert {\cal{P}}^{\prime})
\bar{L}_{2}^{H}(k^{\perp} \vert {\cal{P}}) \nonumber \\ && +
L_{2}^{H^{\prime}}(k^{\perp} \vert {\cal{P}}^{\prime})
\bar{L}_{1}^{H}(k^{\perp} \vert {\cal{P}}) +
L_{2}^{H^{\prime}}(k^{\perp} \vert {\cal{P}}^{\prime})
\bar{L}_{2}^{H}(k^{\perp} \vert {\cal{P}})   \,\,.
\end{eqnarray}
According to (20) we have
\begin{eqnarray*}
 S^{-1}_{i}(k^{\perp})=c^{\nu}_{i}-\hat{k}^{\perp}s^{\nu}_{i}
\end{eqnarray*}
with
\begin{eqnarray*}
c^{\nu}_{i} \equiv cos \nu_{i}(k^{\perp})\,\,,\,\,\,\,\,\,
s^{\nu}_{i} \equiv sin \nu_{i}(k^{\perp}) \,\,.
\end{eqnarray*}
After calculating the trace in (74) one obtains
\begin{eqnarray}
I_{bci}^{(PS)\mu}(v,v^{\prime})&=& i4\pi(v+v^{\prime})^{\mu}
\int \frac{d^{3}k^{\perp}} {(2\pi)^{3}}
(c^{\nu}_{b}c^{\nu}_{c} - s^{\nu}_{b}s^{\nu}_{c})
W(k^{\perp} \vert v,v^{\prime})  \,\,.
\end{eqnarray}
Inserting expression (76) for $I_{bci}^{(PS)\mu}$ into (68) one obtains for
the matrix element in the case of a semileptonic decay of a pseudoscalar $B$
meson into a pseudoscalar meson of the type $(c \bar{i}), \,\,i=u,d,s,c$
within the heavy quark effective theory the result
\begin{eqnarray}
< (l\nu) H^{\prime}_{ji} \vert W^{(3)}_{semi} \vert H_{ib} >
&=& \frac{N_{c}}{3} (2\pi)^{4} \delta({\cal{P}}-{\cal{P}}^{\prime}-
{\cal{P}}_{L}) <l \nu \vert l_{\mu} \vert 0> \nonumber  \\
&& \sqrt{M M^{\prime}} \bigl[ \xi_{+}(v \cdot v^{\prime}) (v+v^{\prime})^{\mu}
+ \xi_{-}(v \cdot v^{\prime}) (v - v^{\prime})^{\mu} \bigr] \,\,\,\,\,
\end{eqnarray}
with the Isgur--Wise functions $[8]$ of the form
\begin{eqnarray}
\xi_{+}(v\cdot v^{\prime}) &=& \frac{1}{\sqrt{M M^{\prime}}}
\frac{1}{(2\pi)^{2} \sqrt{\omega \omega^{\prime}}} \frac{G}{\sqrt{2}} V_{jb}
\nonumber \\
&& \cdot \int \frac{d^{3}k^{\perp}} {(2\pi)^{3}}
(c^{\nu}_{b}c^{\nu}_{c} - s^{\nu}_{b}s^{\nu}_{c})
W(k^{\perp} \vert v,v^{\prime}) \,\,, \\ \nonumber \\
\xi_{-}(v \cdot v^{\prime}) &=& 0 \,\,, \nonumber
\end{eqnarray}
where $W(k^{\perp} \vert v,v^{\prime})$ is defined by (75).

\vspace{0.5cm}

{\large \bf 8. Summary and conclusions }

\vspace{0.5cm}

In this paper we have presented the bilocal field approach for relativistic
covariant potential models in $ QCD $. The main issue consists in the
derivation of general integral expressions for meson properties as decay
constants and semileptonic decay amplitudes. Therefore, it has been necessary
to handle moving bound states. Doing this a special feature of our model has
been important, namely, that the potential kernel (11) moves together with the
bound state because of the presence of the vector $\eta_{\mu}$. Furthermore,
to obtain semileptonic decay amplitudes we have established the relation
between the bilocal field approach for our model and heavy quark effective
theory. In this way we have been able to obtain an integral expression for
the Isgur--Wise function.

The formulas for the meson decay constants (61) and the Isgur--Wise function
(78) appearing in the semileptonic decay amplitude (77) depend on the concrete
form of the potential and contain trigonometric functions and meson wave
functions. Thereby, the trigonometric functions fulfil together with the
energy function the system of equations (23) which has been derived from the
Schwinger--Dyson equation. For constituent quark masses one has the relations
(27) and the system (23) simplifies significantly. The meson wave functions
satisfy systems of equations (50)-(52) resulting from the Bethe--Salpeter
equation. To obtain numerical results for the physical quantities one has to
solve these systems of equations for concrete potentials. Work in this
direction by using different approximations is in progress.

We hope that the hadronization scheme with the use of the Foldy--Wouthuysen
transform can be applied to the hadronization of quarkonia of the type
$ (Q \bar{Q})$ as well. This would allow one to describe also
nonleptonic decays of heavy mesons in the same fashion.

\newpage
\vspace{2cm}
\begin{center}
{\large \bf Tables}
\end{center}
\vspace{2cm}

\begin{center}
\begin{tabular} {|l|l|l|}
\hline
Potential   &                             V(r)   &  V(${\bf p}$) \\
\hline \hline
 & & \\
Coulomb    & $\frac{4}{3} \frac{\alpha_s}{r}  $  &
$ -(\frac{4}{3} \alpha_s) \frac{4 \pi}{ {\bf p}^2  }  $            \\
 & & \\
Linear     & $ a r   $ & $ -a \frac{8 \pi}{ {\bf p}^4 }  $     \\
 & & \\
Oscillator & $ b r^2 $ & $ -b (2\pi)^3 \Delta_{ {\bf p}} \delta ({\bf p}) $ \\
 & & \\
Yukawa     & $ \frac{\alpha }{r} e^{- \eta r}   $ &
$ -\alpha \frac{4\pi}{ {\bf p}^2 + \eta^2  }     $  \\
 & & \\
NJL        & $ V_0 \delta(r)    $  & $ V_0 $ \\
 & & \\
Constant   & $  V_0  $                   & $ V_0 (2\pi)^3 \delta ({\bf p} )$ \\
 & & \\
\hline
\end{tabular}
\vspace{1.5cm}

Tab.1: Interaction potentials in $x$ space and momentum space.
\end{center}

\newpage
\vspace{2cm}
\begin{center}
{\large \bf Figure captions}
\end{center}
\vspace{2cm}
Fig.1: Diagram for the semileptonic decay of a heavy meson
${\cal{M}}(x_{2},x_{3})$ into a heavy meson ${\cal{M}}(x_{6},x_{1})$ and
a leptonic current $\hat{L}(x_{4},x_{5})$ in bilocal field theory.\\
\vspace{1cm} \\
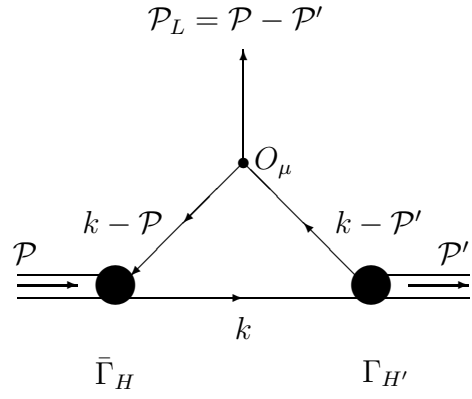
Fig.2: Diagram corresponding to the integral $I^{\mu}_{bji}$, eq.(69),
figurating in the semileptonic decay amplitude (68).

\newpage
\vspace{2cm}
\begin{center}
{\large \bf Figures}
\end{center}
\vspace{2cm}
\unitlength=1mm
\special{em:linewidth 0.4pt}
\linethickness{0.4pt}
\begin{picture}(150.00,127.00)
\put(120.00,108.00){\vector(0,1){15.00}}
\put(90.00,90.00){\vector(1,0){30.00}}
\put(120.00,90.00){\line(1,0){30.00}}
\put(150.00,93.00){\line(-1,0){15.00}}
\put(105.00,93.00){\line(-1,0){15.00}}
\put(135.00,93.00){\vector(-1,1){7.00}}
\put(128.00,100.00){\line(-1,1){8.00}}
\put(120.00,108.00){\vector(-1,-1){8.00}}
\put(112.00,100.00){\vector(-1,-1){7.00}}
\put(137.00,91.67){\circle*{5.20}}
\put(103.00,91.67){\circle*{5.20}}
\put(90.00,91.67){\vector(1,0){8.00}}
\put(142.00,91.67){\vector(1,0){8.00}}
\put(19.00,90.00){\vector(1,0){6.00}}
\put(25.00,90.00){\vector(1,0){25.00}}
\put(50.00,90.00){\vector(1,0){25.00}}
\put(75.00,90.00){\line(1,0){7.00}}
\put(82.00,93.00){\vector(-1,0){7.00}}
\put(75.00,93.00){\line(-1,0){8.00}}
\put(67.00,93.00){\vector(-1,1){7.00}}
\put(60.00,100.00){\line(-1,1){8.00}}
\put(52.00,108.00){\vector(0,1){7.00}}
\put(52.00,115.00){\line(0,1){8.00}}
\put(49.00,123.00){\vector(0,-1){8.00}}
\put(49.00,115.00){\line(0,-1){7.00}}
\put(49.00,108.00){\vector(-1,-1){8.00}}
\put(41.00,100.00){\line(-1,-1){7.00}}
\put(34.00,93.00){\vector(-1,0){9.00}}
\put(25.00,93.00){\line(-1,0){6.00}}
\put(34.00,93.00){\circle*{0.67}}
\put(34.00,93.00){\circle*{1.33}}
\put(34.00,90.00){\circle*{1.33}}
\put(67.00,93.00){\circle*{1.33}}
\put(67.00,90.00){\circle*{1.33}}
\put(49.00,108.00){\circle*{1.33}}
\put(52.00,108.00){\circle*{1.33}}
\put(120.00,108.00){\circle*{1.33}}
\put(67.00,86.00){\makebox(0,0)[cc]{$x_1$}}
\put(34.00,86.00){\makebox(0,0)[cc]{$x_2$}}
\put(34.00,97.00){\makebox(0,0)[cc]{$x_3$}}
\put(67.00,97.00){\makebox(0,0)[cc]{$x_6$}}
\put(44.00,108.00){\makebox(0,0)[cc]{$x_4$}}
\put(56.00,108.00){\makebox(0,0)[cc]{$x_5$}}
\put(50.00,127.00){\makebox(0,0)[cc]{$\hat{L}(x_4,x_5)$}}
\put(26.67,80.00){\makebox(0,0)[cc]{${\cal M}(x_2,x_3)$}}
\put(76.00,80.00){\makebox(0,0)[cc]{${\cal M}(x_6,x_1)$}}
\put(91.00,96.00){\makebox(0,0)[cc]{${\cal P}$}}
\put(148.00,96.00){\makebox(0,0)[cc]{${\cal P}^{\prime}$}}
\put(120.00,86.00){\makebox(0,0)[cc]{$k$}}
\put(138.00,100.00){\makebox(0,0)[cc]{$k-{\cal P}^{\prime}$}}
\put(104.00,100.00){\makebox(0,0)[cc]{$k-{\cal P}$}}
\put(124.00,108.00){\makebox(0,0)[cc]{$O_{\mu}$}}
\put(119.00,127.00){\makebox(0,0)[cc]
    {${\cal P}_{L}={\cal P} - {\cal P}^{\prime}$}}
\put(103.00,80.00){\makebox(0,0)[cc]{$\bar{\Gamma}_H$}}
\put(139.00,80.00){\makebox(0,0)[cc]{$\Gamma_{H^{\prime}}$}}
\put(50.00,65.00){\makebox(0,0)[cc]{${\bf Figure \,\,1.}$}}
\put(120.00,65.00){\makebox(0,0)[cc]{${\bf Figure \,\,2.}$}}
\end{picture}

\newpage

\begin{thebibliography}{99}


\bibitem{1.} For a recent review see: A. Ali, {\em{B-Decays--Introduction}},
             DESY-Preprint 91-137 (to be published in {\em{B Decays}},
             ed. S. Stone (World Scientific Publishers, Singapore).
\bibitem{2.} H. Kleinert, {\em{Phys. Lett.}} {\bf{B62}} (1976) 429 and
             in: {\em{Understanding the Fundamental Constituents
             of Matter}} (Erice Lectures 1976), ed. A. Zichichi
             (Plenum, New York, 1978).
\bibitem{3.} D. Ebert and V.N. Pervushin, {\em {Theor. Math. Phys.}} {\bf{36}}
	     (1978) 759;\\
             D. Ebert, H. Reinhardt and V.N. Pervushin, {\em{Sov. J. Part.
             Nucl.}} {\bf{10}} (1979) 444.
\bibitem{4.} Yu.L. Kalinovsky et al., {\em{Sov. J. Nucl. Phys.}}
             {\bf{49}} (1989) 1059; \\
             V.N. Pervushin, {\em{Nucl. Phys.}} {\bf{B15}} (Proc. Suppl.)
             (1990) 197.
\bibitem{5.} V.N. Pervushin et al., {\em{Fortschr. Phys.}} {\bf{38}} (1990)
             333; \\
             Yu.L. Kalinovsky, L. Kaschluhn and V.N. Pervushin,
             {\em{Fortschr. Phys.}} {\bf{38}} (1990) 353; \\
             Yu.L. Kalinovsky et al., {\em{Few--Body Systems}} {\bf{10}}
             (1991) 87.
\bibitem{6.} Yu.L. Kalinovsky, L. Kaschluhn and V.N. Pervushin,
             {\em{Phys. Lett.}} {\bf{231}} (1989) 288.
\bibitem{7.} A. Le Yaouanc et al., {\em{Phys. Rev.}} {\bf{31}} (1985) 137
\bibitem{8.} N. Isgur and M.B. Wise, {\em{Phys. Lett.}} {\bf{B232}}
             (1989) 113; {\em{ibid}} {\bf{237}} (1990) 527.
\bibitem{9.} E. Eichten and B. Hill, {\em {Phys. Lett}} {\bf{B234}}
             (1990) 511; \\
             H. Georgi, {\em {Phys. Lett}} {\bf{B240}}
             (1990) 447; \\
             B. Grinstein, {\em {Nucl. Phys.}} {\bf{B339}}
             (1990) 253.
\bibitem{10.} For recent reviews see: \\
             M.B. Wise, {\em{New Symmetries of the Strong Interaction}}, Lake
             Louise Winter Inst., Lake Louise, Canada, Feb. 1991,
             CALT-68-1721; \\
             H. Georgi, {\em{Heavy Quark Effective Field Theory}},
             HUPT-91-a039; \\
             B. Grinstein, {\em{Lectures on Heavy Quark Effective Theory}},
             Workshop on High Energy Phenomenology, Mexico City, Mexico, July
             1991, HUPT-91-A040 and SSCL-Preprint 17.
\bibitem{11.} R.T. Cahill, J. Praschifka and C.J. Burden, {\em{Aust. J. Phys.}}
	     {\bf{42}} (1989) 161.
\bibitem{12.} V.N. Pervushin, {\em{Riv. Nuovo Cim.}} {\bf{8}} (1985) 1; \\
             Nguyen Suan Han and V.N. Pervushin, {\em{Fortschr. Phys.}}
             {\bf{37}} (1986) 611 and {\em{Mod. Phys. Lett.}} {\bf{A2}}
             (1987) 367.

\end{thebibliography}
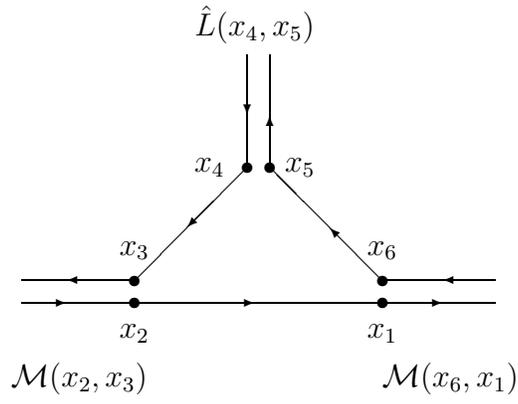
\end{document}